\documentclass[journal, letterpaper, twocolumn]{IEEEtran}
\usepackage[english]{babel}
\usepackage[utf8x]{inputenc}
\usepackage{amsmath}
\usepackage{graphicx}
\usepackage{array}
\usepackage{makecell}
\newcolumntype{x}[1]{>{\centering\arraybackslash}p{#1}}
\usepackage{tikz}
\newcommand\diag[4]{%
  \multicolumn{1}{|p{#2}|}{\hskip-\tabcolsep
  $\vcenter{\begin{tikzpicture}[baseline=0,anchor=south west,inner sep=#1]
  \path[use as bounding box] (0,0) rectangle (#2+2\tabcolsep,\baselineskip);
  \node[minimum width={#2+2\tabcolsep-\pgflinewidth},
        minimum  height=\baselineskip+\extrarowheight-\pgflinewidth] (box) {};
  \draw[line cap=round] (box.north west) -- (box.south east);
  \node[anchor=south west] at (box.south west) {#3};
  \node[anchor=north east] at (box.north east) {#4};
 \end{tikzpicture}}$\hskip-\tabcolsep}}
\usepackage{subcaption}
\usepackage{cite}
\usepackage{multirow}

\usepackage{algorithm}
\usepackage{algpseudocode}
\usepackage{booktabs}
\usepackage{url}
\usepackage{mathtools}
\newcommand{\comment}[1]{ }
\newcommand{\contentset}{{\cal F}}
\newcommand{\PLFU}{{{pLFU}}}
\newcommand{\ULFU}{{{sLFU}}}
\newcommand{\estULFU}{{{$\widehat{s}$LFU}}}
\newcommand{\Neigh}{{{\cal {N}}}}
\newcommand{\util}{{{\cal {S}}}}

\newcommand{\network}{{{\cal {S}}_{neigh}}}
\newcommand{\node}{{{\cal {S}}_{local}}}

\title{Responsive Content-Centric Delivery in Large Urban Communication Networks: A LinkNYC Use-Case}

\author{Hassan Sinky, Bassem Khalfi, Bechir Hamdaoui, and Ammar Rayes$^{\dag}$
\thanks{This work was supported in part by the US National Science Foundation (NSF) under NSF award CNS-1162296.}
~\\
\small Oregon State University, \small Corvallis, OR 97331,
\small sinkyh,khalfib,hamdaoui@oregonstate.edu~\\
$^{\dag}$ \small Cisco Systems, San Jose, CA 95134, rayes@cisco.com
\vspace{-0.1in}
}

\IEEEoverridecommandlockouts
\begin{document}
\maketitle

\begin{abstract}
Large urban communication networks such as smart cities are an ecosystem of devices and services cooperating to address multiple issues that greatly benefit end users, cities and the environment. LinkNYC is a first-of-its-kind urban communications network aiming to replace all payphones in the five boroughs of New York City (NYC) with kiosk-like structures providing free public Wi-Fi. We consolidate these networks with standalone edge cloud devices known as cloudlets and introduce geographically distributed content delivery cloudlets (CDCs) to store popular Internet content closer to end users; essential in environments with diverse and dynamic content interests. A content-centric and delivery framework is proposed leveraging NYC's population densities and CDCs for interest-based in-network caching. Analysis shows that although the adoption of multiple CDCs dramatically improves overall network performance, advanced caching policies are needed when considering increased content heterogeneity. Thus, we propose popularity-driven (\PLFU) and cooperation-based (\ULFU) caching policies at individual CDCs to account for user and content dynamics over time. The amalgamation of urban population densities, multiple CDC placements and smarter caching techniques helps exploit the ultimate benefits of a content-centric urban communications network and dramatically improves overall network performance and responsiveness. Our proposed solutions are validated using LinkNYC as a use-case.

\end{abstract}

\section{Introduction}
\label{sec:p6-intro}
The proliferation of Internet devices has resulted in efforts to integrate various wireless access technologies for improved performance, increased services and inter-connectivity of end users. The recent growth in data demand has prompted researchers to come up with new wireless techniques (e.g., MIMO~\cite{hamdaoui2007cross-b,hamdaoui2007characterization}, cooperative communication~\cite{su2008cooperative}, femtocells~\cite{chakchouk2012uplink}, etc.) and develop new technologies (e.g., cognitive radio~\cite{venkatraman2010opportunistic}, LTE~\cite{dechene2014energy}, etc.) to be able to meet this high demand. This integration allows for large geographic locations to be serviced providing millions of end users with continuous connectivity and optimal quality of experience (QoE). However, the world has seen unprecedented urban population growth over the years. In fact, the number of urban residents has increased by nearly 60 million a year. By 2050, it is estimated that 70\% of the world's population will be living in cities\footnote{World population data sheet: \url{http://www.prb.org/}}. Urban communication networks and content delivery networks have been introduced to leverage these technologies to better service cities and users alike. Content delivery networks are designed to improve overall network performance by bringing data closer to the geographical locations of users.

Urban communication networks have evolved over the years to address urban challenges through the use of information, communication technology and the Internet. Building such a network infrastructure capable of adequately servicing urban locations has become increasingly difficult due to the sheer number of Internet devices and users (e.g., rapidly increasing numbers of Internet of Things (IoT) devices). Research shows that global mobile data traffic will increase sevenfold reaching 49 exabytes per month by 2021, most of which will be mobile video content, with a percentage projected to reach up to 78\% by 2021~\cite{index2017global}. Thus, not only pushing content closer to the end user but also smarter content placement and timely delivery of content are of the utmost importance.

Traditionally, content delivery nodes or datacenters are geographically distributed throughout the world servicing different regions. In large urban networks the same content may be requested by multiple users resulting in the content traversing the network multiple times, over large distances, to and from a remote content delivery node hosting the content. As detailed in Section~\ref{sec:p6-linknyc}, we leverage the LinkNYC infrastructure, a large urban communications network currently underway providing free public Wi-Fi to New York City (NYC) residents, to help propose and design techniques for efficient content delivery in large urban networks. Thus, the focus of this work is twofold. First, we propose and analyze the placement of multiple content delivery cloudlets (CDC), based on user population densities within a large urban communications network, to bring content closer to consumers. Second, smarter CDC storage techniques are proposed to leverage content-centric in-network caching principles based on content popularity for efficient and timely deliveries. Contributions of this work are as follows:

\begin{itemize}
\item Performance analysis of the proposed shift of LinkNYC's infrastructure from a traditional communications network to a content-centric network.
\item Designs and evaluates a CDC placement heuristic which exploits NYC population densities acquired from the 2012 LandScan\textsuperscript{TM} dataset.
\item Establishes that large urban communication networks vastly improve with not only a content-centric shift but also the placement of multiple content delivery cloudlets geographically distributed throughout the network.
\item Designs and evaluates CDC storage techniques for timely and responsive content delivery within large networks.
\item To our knowledge, this work is the first to analyze and leverage LinkNYC's urban communications network through content delivery cloudlets improving overall network performance and stability as well as mobile content delivery and service continuity.
\end{itemize}

The rest of this paper is organized as follows. In Section~\ref{sec:p6-linknyc} content-centric urban communication networks and the LinkNYC infrastructure are introduced, constructed and analyzed. Next, a CDC placement heuristic is proposed in Section~\ref{sec:p6-cdc}. Section~\ref{sec:p6-rtop} introduces potential content storage solutions for responsive content delivery. Proposed solutions are then analyzed and validated using LinkNYC in Section~\ref{sec:p6-analysis}. Finally, the article is concluded in Section~\ref{sec:p6-conclusion}.

\section{Content-Centric Urban Communication Networks}
\label{sec:p6-linknyc}

Coupling urban communication networks with content-centric and delivery principles greatly benefit content producers, consumers and the cities they reside in. Improving their infrastructure using practical approaches to provide more reliable and responsive communications can assist in the technology's overall success. The underlying concept behind content-centric communication networks is to allow a consumer to focus on the desired named content rather than referencing the physical location or named hosts (IP) where that content is stored~\cite{ahlgren2012survey,fayazbakhsh2013less,xylomenos2014survey}. This shift is a product of empirical research resulting in the fact that the vast majority of Internet usage involves data being disseminated from a source to multiple users. The potential benefits of a content-centric adoption include in-network caching to reduce congestion, improved delivery speeds, simpler network configuration and network security at the data level~\cite{NFD:14}. This paper combines content-centric and content delivery principles to improve the performance and reliability of urban networks such as LinkNYC.

\subsection{Related Works}
\label{lab:p6-related}

In \cite{rossi2011caching,retal2017content,bernardini2013mpc,zhang2015survey,ioannou2016survey} content caching in content-centric networks (CCN) is considered to exploit in-network caching and improve end user's QoE. Different caching policies have been proposed which can be classified into three groups: autonomous, centralized, and cooperative~\cite{ioannou2016survey}.
Some research focuses on exploiting popularity to enhance caching in CCN with the aim of reducing the amount of cached items while maintaining a high cache hit rate~\cite{bernardini2013mpc,thar2015efficient}.
Other works~\cite{taleb2015user,bagaa2014service} focus on virtual network function placement approaches while accounting for mobility~\cite{taleb2015user} and service~\cite{bagaa2014service} aspects.
In \cite{bernardini2014socially,yanqing2016socially} social information is exploited by identifying influential users and proactively caching their content in the network.
Combining popularity with cooperation has also been considered~\cite{ming2014age,wang2013collaborative,wang2013intra,zhang2015survey,li2012popularity}. However, a major issue with these approaches is the communication overhead due to added information exchange and signaling, especially in large networks, as well as requiring many network parameters to be optimal. For this reason autonomous caching techniques are often favored. Our work proposes a framework that leverages (i) content delivery techniques to bring content closer to users within large urban communication networks and (ii) content-centric principles for cooperative in-network caching to improve overall network performance and user QoE.

\subsection{Content Delivery Networks}
\label{lab:p6-cdn}

\begin{figure}[]
\centering
\includegraphics[width=.5\textwidth]{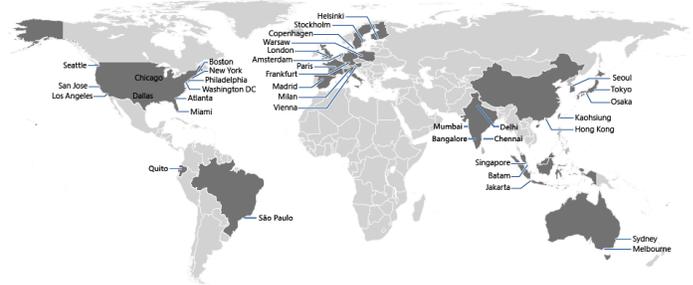}
\caption{\label{fig:azure}Microsoft Azure CDN point of presence locations}
\end{figure}

The principle behind content delivery networks (CDN) is to bring data as close to the geographic location of the user as possible to improve overall network performance. This helps eliminate the need to traverse the Internet for content which reduces infrastructure and bandwidth costs while improving network robustness and quality of experience (QoE). For instance, Microsoft Azure's content delivery network consists of 36 points of presence locations\footnote{Figure~\ref{fig:azure} provided by Microsoft Azure: \url{https://azure.microsoft.com/en-us/documentation/articles/cdn-pop-locations}} distributed throughout the world as shown in Figure~\ref{fig:azure}. However, in addition to being geographically limited, CDN nodes are generally placed at the Internet edges over multiple backbones servicing different regions remotely. Although the concept of bringing content closer to the consumer through caching copies in various geographic locations improves overall network performance; mobile users in large urban networks naturally endure additional latency due to increased mobility, congestion and hops traversed within the network. This can be improved by placing content even closer to the requesting consumer through content-centric networking and delivery principles.

\subsection{Content-Centric Networks}
\label{lab:p6-ccn}

Different Internet architectures have been introduced for content-centric networking which shifts from the standard IP data packets to Named-Data Networking (NDN). NDN is a future Internet architecture focusing on a content-centric Internet as opposed to today's host-centric network architecture~\cite{NFD:14,367,399}. These architectures rely on named-content within the Internet to route and direct the flow of data within a network. In these architectures the content is the focus rather than the physical location where the content is stored. NDN consists of content consumers and producers. Consumers generate interest requests for specific content chunks whereas producers respond to interest requests with data packets. Every node is equipped with a Content Store (CS), Pending Interest Table (PIT), and Forwarding Information Base (FIB). When an interest packet is received it is first placed into the PIT and the CS is checked for data correlating to this interest. If there is a match the interest request is discarded and the corresponding data packet from the CS is returned. Otherwise the interest packet is forwarded based on information in the FIB. Incoming data packets corresponding to pending interests in the PIT are stored in the node's CS. Otherwise, the data packets are dropped. These concepts are used in designing our responsive content delivery framework for large urban networks. Before presenting our solutions, we first introduce and construct the LinkNYC network which is the basis behind our design choices.

\subsection{LinkNYC Framework}
\label{sec:p6-framework}

In November 2014, LinkNYC announced a project plan to provide a first-of-its-kind communications network offering super fast free gigabit Wi-Fi to everyone in New York City (NYC) through the replacement of thousands of payphones with kiosk-like structures called \textbf{Links}. Once completed, LinkNYC will be the largest and fastest free public Wi-Fi network in the world. The \textbf{Links} are designed as an update to the standard phone booth and act as Wi-Fi hotspots while also providing basic services such as advertisements, free phone calls, device charging, touchscreen for Internet browsing to access city services, maps and directions. Revenue generated by the \textbf{Links}, through kiosk Ads that are displayed on 55 inch displays, is used to maintain the LinkNYC infrastructure. Each Link is equipped with 802.11ac Wi-Fi technology yielding real world download and upload speeds of 300 and 320 Mbps respectively with an average latency of 5 ms and coverage area of up to 45 meters depending on location~\cite{engadget:linknyc:2016}. This promising self-contained urban communications network provides cities with revenue and analytics while offering consumers free, continuous and reliable connectivity. Our proposed framework is designed for large urban communication networks and tested on a virtual construction of the LinkNYC network using their payphone locations, NYC population densities and hardware expectations based on specifications provided to the public by LinkNYC.

\begin{figure}[]
\centering
\includegraphics[trim={120 13 0 13}, clip,width=.4\textwidth]{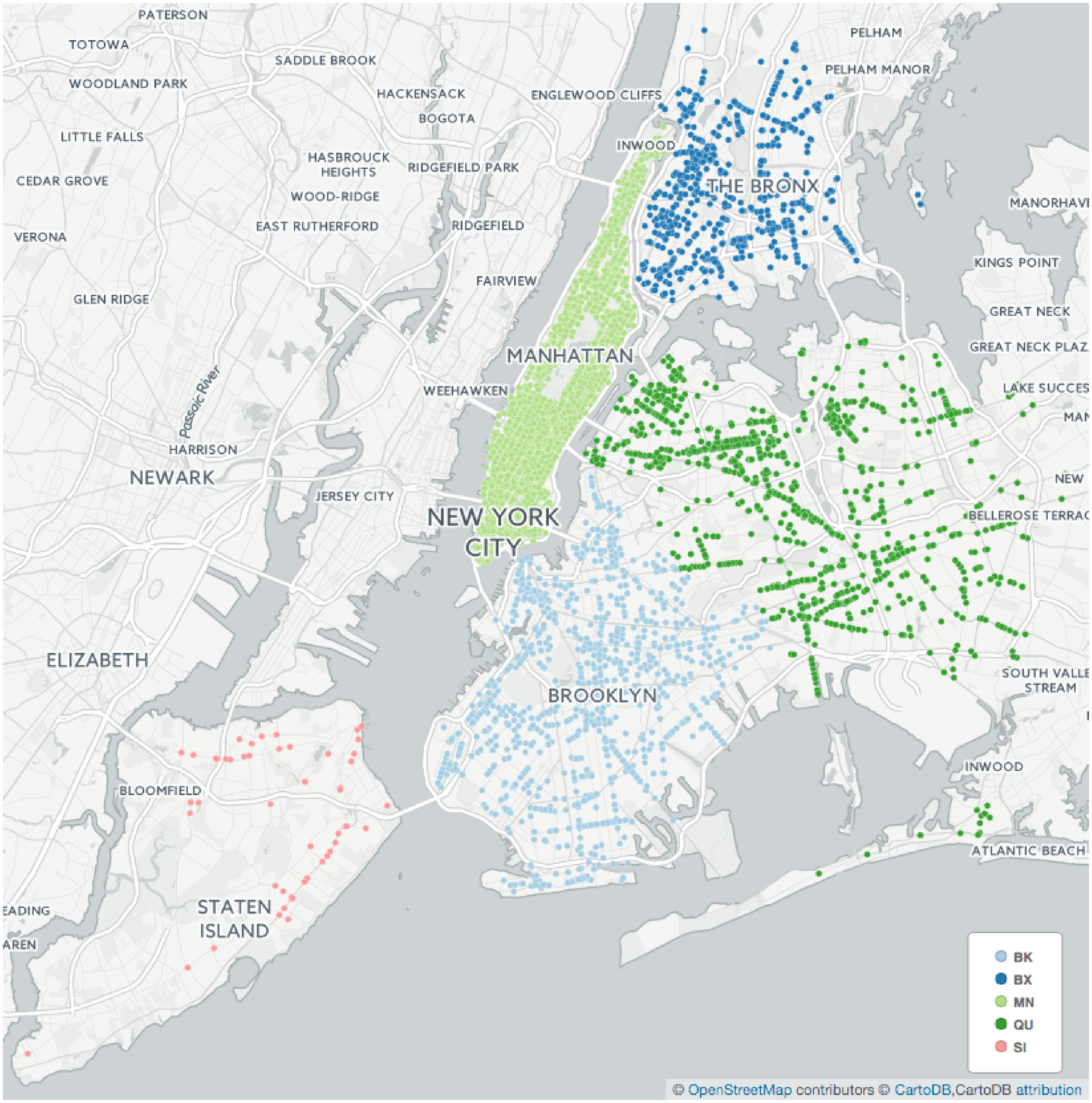}
\caption{\label{fig:nyc}Geographic locations of payphones in NYC}
\end{figure}

\subsection{Borough Analysis and Topology Construction}

Naturally, the number of potential CDCs depends on the number of currently installed payphone locations\footnote{NYC Open Data: \url{https://nycopendata.socrata.com }} summarized in Table~\ref{tab:nyc-summary}. Manhattan is the most dense of the five boroughs with 3,409 payphones and an average distance between them of 43 meters. On the other hand, Staten Island is the most sparse with only 51 payphones and an average distance of 606 meters.

LinkNYC cloudlets are categorized based on their borough as shown in Figure~\ref{fig:nyc}. Since the connectivity of NYC's payphone backhaul is unknown, we assume that cloudlets are physically connected (e.g., by fiber optic cables) to their nearest neighbors. Given a particular NYC borough (e.g. Brooklyn), we construct an Euclidean minimum spanning tree (EMST) as shown in Figure~\ref{fig:bk_cl} using Prim's algorithm~\cite{gower1969minimum,manen2013prime} where edge weights are equal to the geographic distance between cloudlets. EMSTs are useful for telecommunications companies to decide for e.g. where to deploy fiber optic cables considering longer cables are more costly. This results in a network topology with $N-1$ edges where $N$ is the number of cloudlets in a particular borough. This constructed LinkNYC infrastructure will be used later to validate our proposed solutions.

\begin{figure*}
        \centering
        \begin{subfigure}[c]{.15\textwidth}
                \includegraphics[width=\linewidth]{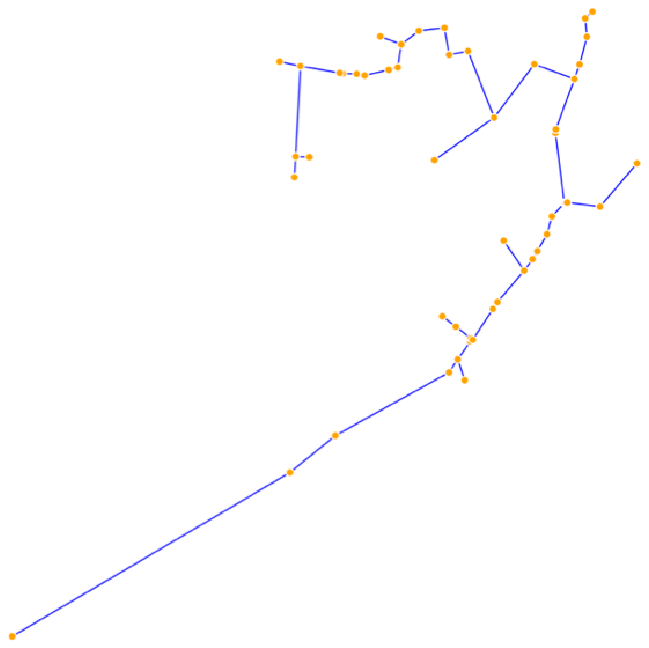}
                \caption{Staten Island}
                \label{fig:si}
		\end{subfigure}\begin{subfigure}[c]{.15\textwidth}
                \includegraphics[width=\linewidth]{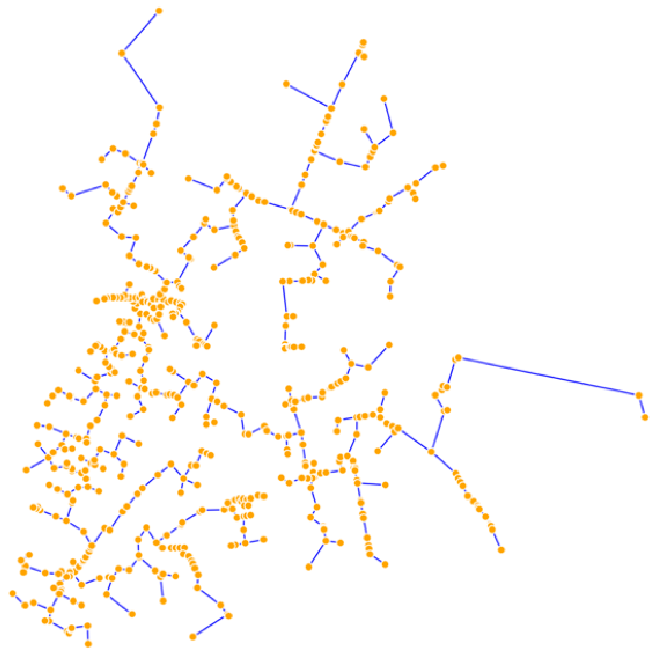}
                \caption{Bronx}
                \label{fig:bx}
		\end{subfigure}\begin{subfigure}[c]{.15\textwidth}
                \includegraphics[width=\linewidth]{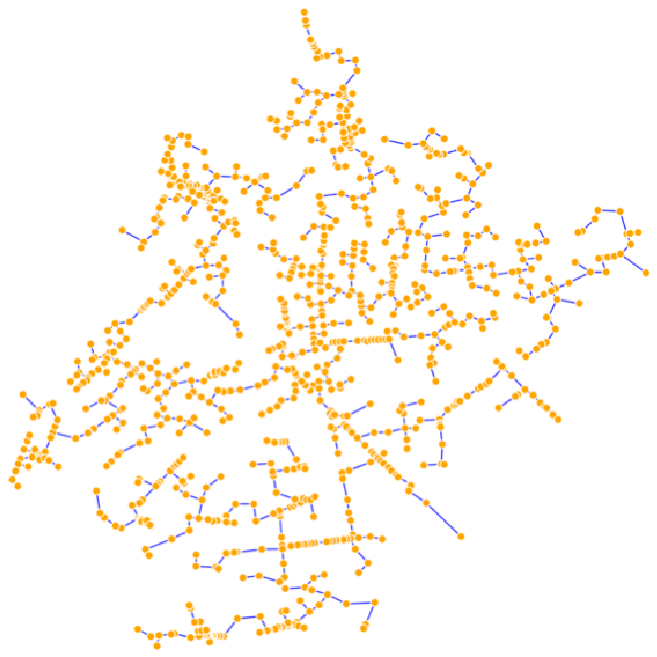}
                \caption{Brooklyn}
                \label{fig:bk}
		\end{subfigure}\begin{subfigure}[c]{.15\textwidth}
                \includegraphics[width=\linewidth]{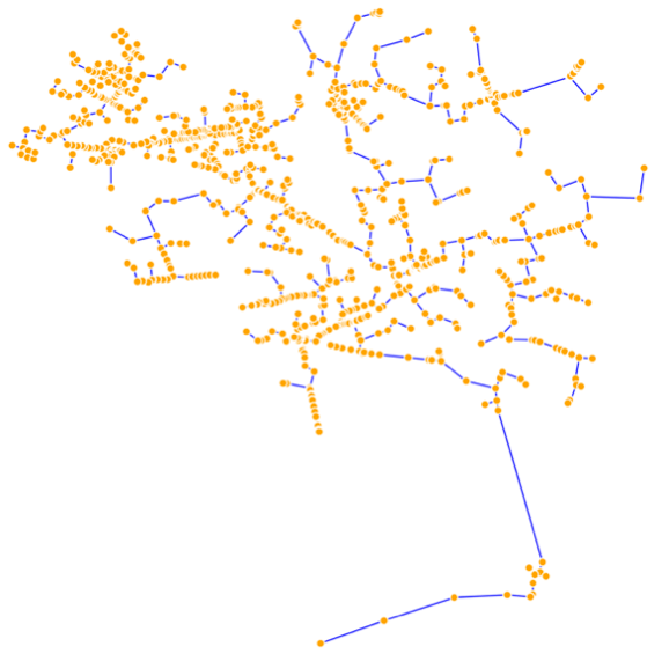}
                \caption{Queens}
                \label{fig:qu}
		\end{subfigure}\begin{subfigure}[c]{.15\textwidth}
                \includegraphics[width=\linewidth]{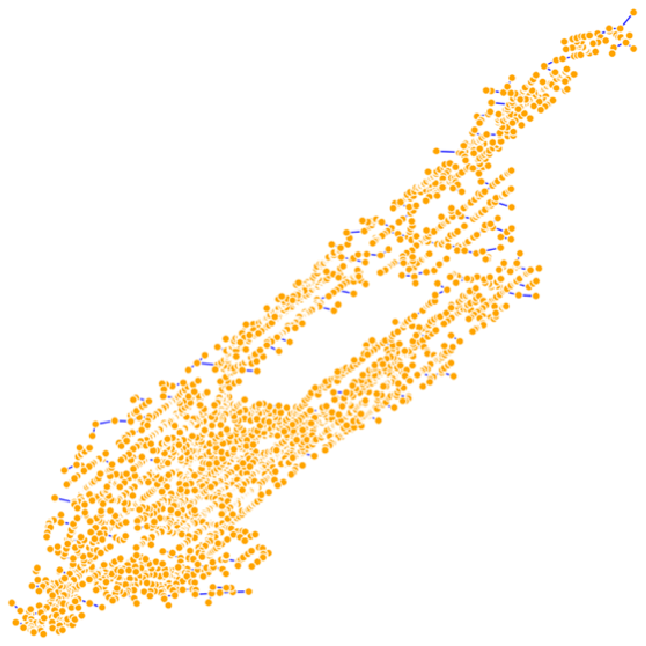}
                \caption{Manhattan}
                \label{fig:mn}
		\end{subfigure}
        \caption{LinkNYC network construction using minimum spanning trees}
        \label{fig:emst}
\end{figure*}

In order to leverage content-centric and content delivery networking with large urban communications networks such as LinkNYC's infrastructure, this work proposes an architectural addition of edge cloud devices referred to as content-delivery cloudlets (CDCs) to better service end users. CDCs are small-scale cloud datacenters that aim to bring data closer to mobile users and are typically located near the edge of the Internet~\cite{cloudlet,chen2016privacy,jiang2015energy,chen2015emc,xu2013survey}. Upgrading, for example, LinkNYC's \textbf{Links} with these CDCs offers a different dimension to content delivery networking. Unlike traditional content delivery networks, where a limited number of remote servers or nodes are distributed throughout the world, the proposed approach selects and designates a subset of LinkNYC's \textbf{links} as CDCs (i.e., producers of content). These selected CDCs are each equipped with an $L2$ storage cache that is much larger than the $L1$ storage cache available on other cloudlets. Content is locally cached as it traverses the network to its destination. For this paper, we assume storage capabilities are limited to CDCs only (i.e. $L2$ cache only) as to avoid excessive deployment costs. In order to decide which CDCs will be selected for providing content delivery services a hierarchical clustering technique is applied to the borough topologies based on NYC's population density distribution detailed in the following section.

\section{Content Delivery Cloudlet Selection and Placement}
\label{sec:p6-cdc}

As consumers become increasingly mobile with dynamic content interests the placement of CDCs in large cities becomes both crucial and challenging. Specifically, mobile users that encounter multiple cloudlets across a path and undergo frequent handoffs experience service quality continuity issues~\cite{hassan:15,sinky2013cross}. With added dynamic content interests, if CDCs' placement is not carefully designed, these issues may be exacerbated. Relying on a single CDC is insufficient to meet the demand of dynamic consumers, and thus, having content readily available in multiple nearby CDCs is indispensable to ensure responsive content delivery and maintain good QoE. This avoids the additional costs of traversing the entire network or querying the original publisher for the content. In what follows, we investigate clustering approaches to efficiently select and decide on the placement of multiple CDCs to enable content-centric networking and delivery in smart cities while considering the LinkNYC network as our use-case for applying and evaluating such approaches.

\subsection{Population Distribution}
\label{sec:p6-pop}

\begin{figure}[H]
\centering
\includegraphics[width=.5\textwidth]{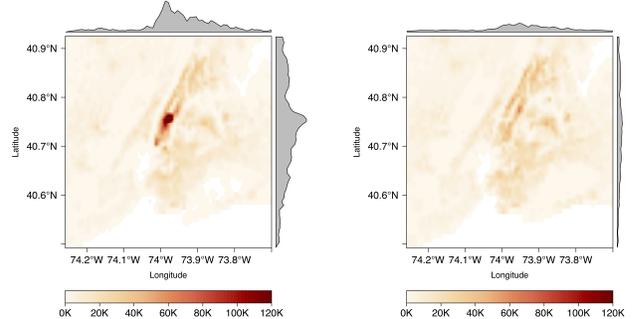}
\caption{\label{fig:linknyc-pop}Daytime versus nighttime population densities}
\end{figure}

In order to reflect realistic applications in practice a complete content-centric LinkNYC framework is designed based on NYC's population density distribution using the United States Department of Energy's ORNL (Oak Ridge National Laboratory) LandScan\textsuperscript{TM} 2012 dataset. LandScan\textsuperscript{TM} uses spatial data, satellite imagery analysis and a multi-variable dasymetric modeling approach to disaggregate census counts to provide accurate ambient population per square kilometer over a 24 hour period. This results in an average population density during the daytime. Nighttime population densities are acquired using the Socioeconomic Data and Applications Center (SEDAC) Gridded Population of the World (GPW) dataset which uses census and satellite data. Average population densities of LinkNYC are illustrated in Figure~\ref{fig:linknyc-pop}; Note, NYC experiences an influx of 4.5 million people during the day which is mostly concentrated in the Manhattan borough. The datasets were then disaggregated to obtain constant population density estimates immediately surrounding individual LinkNYC cloudlets. Given the population estimates a clustering heuristic is applied to each borough for CDC placement.

\subsection{CDC Hierarchical Clustering Heuristic}
\label{sec:p6-clustering}

Cluster analysis is an NP-hard problem, thus, efficient heuristic algorithms are commonly employed that converge to a local optimum depending on the application. For this purpose, we propose a population density based clustering heuristic for large urban communication networks and CDC placement. Once CDC placement is decided it is assumed to be static. That is, CDCs are upgraded \textbf{Links} with enhanced hardware. Thus, frequently changing CDC locations is not practical in a dynamic environment.

Initially, cloudlets are part of the same membership and form a single community. In order to introduce content delivery services to the borough network, a cloudlet is chosen as a CDC based on its average hop count to the remaining cloudlets within its respective community. The probability that requests are initiated from each cloudlet, $i$, is assumed to be proportional to its respective surrounding population density, $\gamma_i$, and is defined as
$r_i=\frac{\gamma_i}{\sum_{j=1}^{N_l} \gamma_j}$, where $N_l$ is the number of Links in the entire borough network (e.g., Brooklyn). The population densities of all LinkNYC's Links---$\gamma_i$ for Link $i$---are estimated as described in Section~\ref{sec:p6-pop}. Let's now denote by $\boldsymbol{S}$ the shortest path matrix that contains the length of the shortest path to and from each Link in the borough network.
Given the cloudlet request probability vector, $\boldsymbol{r}=(r_1,r_2,\ldots,r_{N_l})$, a weighted average shortest path vector,  $\bar{\boldsymbol{s}}=\boldsymbol{S}\cdot\boldsymbol{r}$, is computed per community and whose entry values represent the weighted average hop count provided by each cloudlet to and from each other. Then, the cloudlet with the minimum sum of weighted average hop counts to remaining cloudlets is selected as the CDC, i.e., $\arg\min\bar{\boldsymbol{s}}$.
This ensures content is placed as close to the geographic location of potential consumers within a community as possible.
Once selected, the incident edge between the CDC and the cloudlet with the minimum average hop count is removed forming two disjoint communities.
Then, for each community, a single cloudlet is selected as a CDC by recomputing vector $\bar{\boldsymbol{s}}$.
This results in a set of $\bar{\boldsymbol{s}}$-vectors where the community experiencing a higher average hop count is chosen for additional CDC placement, i.e., $\arg\max\left(\min\bar{\boldsymbol{s}}^{(1)},\min\bar{\boldsymbol{s}}^{(2)},...,\min\bar{\boldsymbol{s}}^{(n)}\right)$ where $n$ represents the current number of communities.
Edges are progressively removed from the original topology until no edges remain, resulting thus in a set of communities equal to the number of cloudlets.

\begin{table}[h]
\begin{center}
\begin{tabular}{llll}
\toprule
\textbf{Borough} & \textbf{\# Payphones} & \textbf{Avg. distance} & \textbf{CDCs} \\
\midrule
	Manhattan   & 3409 & 43.2 m & 50\\
    Queens & 1042 & 136.8 m & 25 \\
    Brooklyn & 1004 & 150.8 m & 25 \\
	Bronx & 591 & 125.5 m & 20 \\
    Staten Island & 51 & 606 m & 10\\
    Total & 6097 & 212.5 m & 130 \\
	\bottomrule
\end{tabular}
\caption{Payphones in the five boroughs of NYC}
\label{tab:nyc-summary}
\end{center}
\end{table}

\begin{figure}[h]
        \centering
         \begin{subfigure}[c]{.45\textwidth}
                \includegraphics[width=\linewidth]{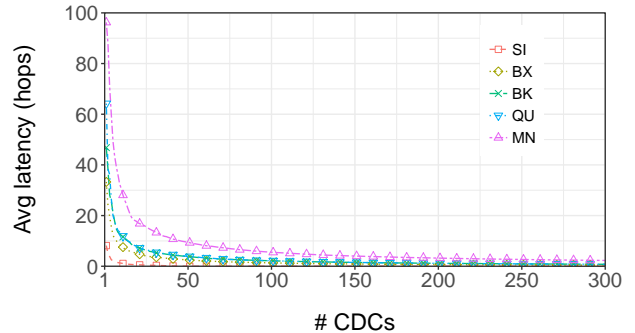}
                \caption{Average hop count as a function of CDCs}
                \label{fig:cdn_elbow}
		\end{subfigure}
        \begin{subfigure}[c]{.41\textwidth}
                \includegraphics[trim={0 60 0 60},clip,width=\linewidth]{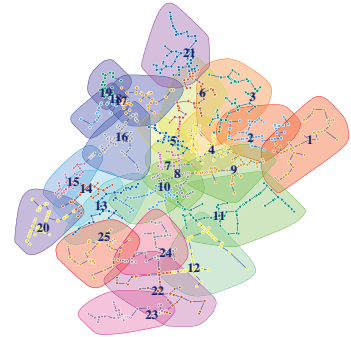}
                \caption{Clustering of Brooklyn's LinkNYC network}
                \label{fig:bk_cl}
        \end{subfigure}
        \caption{Borough latency and CDC placement}
        \label{fig:cdn_placement}
\end{figure}

The intuition behind this heuristic is to assign more CDCs to highly populated areas. Traditionally, large urban networks rely on a regional CDN (i.e., CDCs = 0) for content caching which provides inadequate responsiveness as content would not only still need to be fetched from the original publisher but also traverse the large urban network incurring additional latencies.
Incorporating in-network caching through the addition of a single CDC immediately improves latency by storing content closer to users. However, we can see from Figure~\ref{fig:cdn_elbow} that a single CDC still does not facilitate timely content delivery as the average latency to access the CDC is high. To determine an adequate number and initial placement of borough CDCs during the daytime, the elbow in the CDC curve in Figure~\ref{fig:cdn_elbow} is estimated and used to decide on the number of CDCs to be chosen and are summarized in Table~\ref{tab:nyc-summary}. Figure~\ref{fig:bk_cl} illustrates the resulting communities and CDCs for Brooklyn's topology (i.e., CDCs = 25).

This analysis shows that large urban networks, such as LinkNYC, can benefit greatly with not only the adoption of a content-centric infrastructure (i.e., single CDC) but also with the incorporation of multiple content delivery cloudlets within its urban communications network. Overall a cluster-based placement approach exhibits promising and more stable results yielding lower hop counts and achieving on average a higher rate of cache hits which in turn improves content delivery speeds. Next, we propose content popularity-driven and cooperative caching techniques deployed by CDCs for improved infrastructure performance.

\section{Popularity-Driven and Cooperative Content-Centric Caching}
\label{sec:p6-rtop}

Internet content of interest to users far surpasses the storage capacity of the CDCs, thereby resulting in frequent content requests from the original content publisher. When content is not stored at intermediate cloudlets, it must be fetched from a remote datacenter incurring additional latency and service continuity issues~\cite{sinky2016cloudlet}. Naturally, by increasing cache sizes or the number of CDCs, cloudlets will be able to store and push content closer to end users, but this cannot be done without additional hardware and cache deployment costs. Depending on their needs and resource availability, city officials, policy makers, and network administrators manage to find the balance between acceptable QoE and deployment costs. In this work, we opt for techniques that can improve network performance and QoE while avoiding hardware costs. Hereinafter \textit{cloudlet} and \textit{CDC} will be used interchangeably.

\subsection{Popularity-Driven Content Caching}
\label{sec:p6-plfu}
Traditionally, caching consists of fetching content upon request and storing it locally based on some cache replacement technique, such as First-In-First-Out (FIFO), Random Replacement (RR), Most Recently Used (MRU), Least Recently Used (LRU), Least Frequently Used (LFU), etc~\cite{al2004performance}, in anticipation of future content requests. Current cache information (i.e., content, timestamps, queue, etc) is then used to make replacement decisions (i.e., evicting least used content). LRU is among the most traditional reactive cache replacement policies and is still used in practice today~\cite{lee2001lrfu}. These traditional solutions, however, are not suitable for Internet content delivery in highly populated cities, such as NYC, mainly due to the diversity, heterogeneity, volume, and dynamic nature of Internet content. In what follows from this section, we propose a content-centric LFU caching method suitable for these highly populated cities that incorporates and relies on the popularity of content encountered by CDCs to decide on which content to be cached.

\begin{figure*}
        \centering
        \begin{subfigure}[c]{.4\textwidth}
                \includegraphics[width=\linewidth]{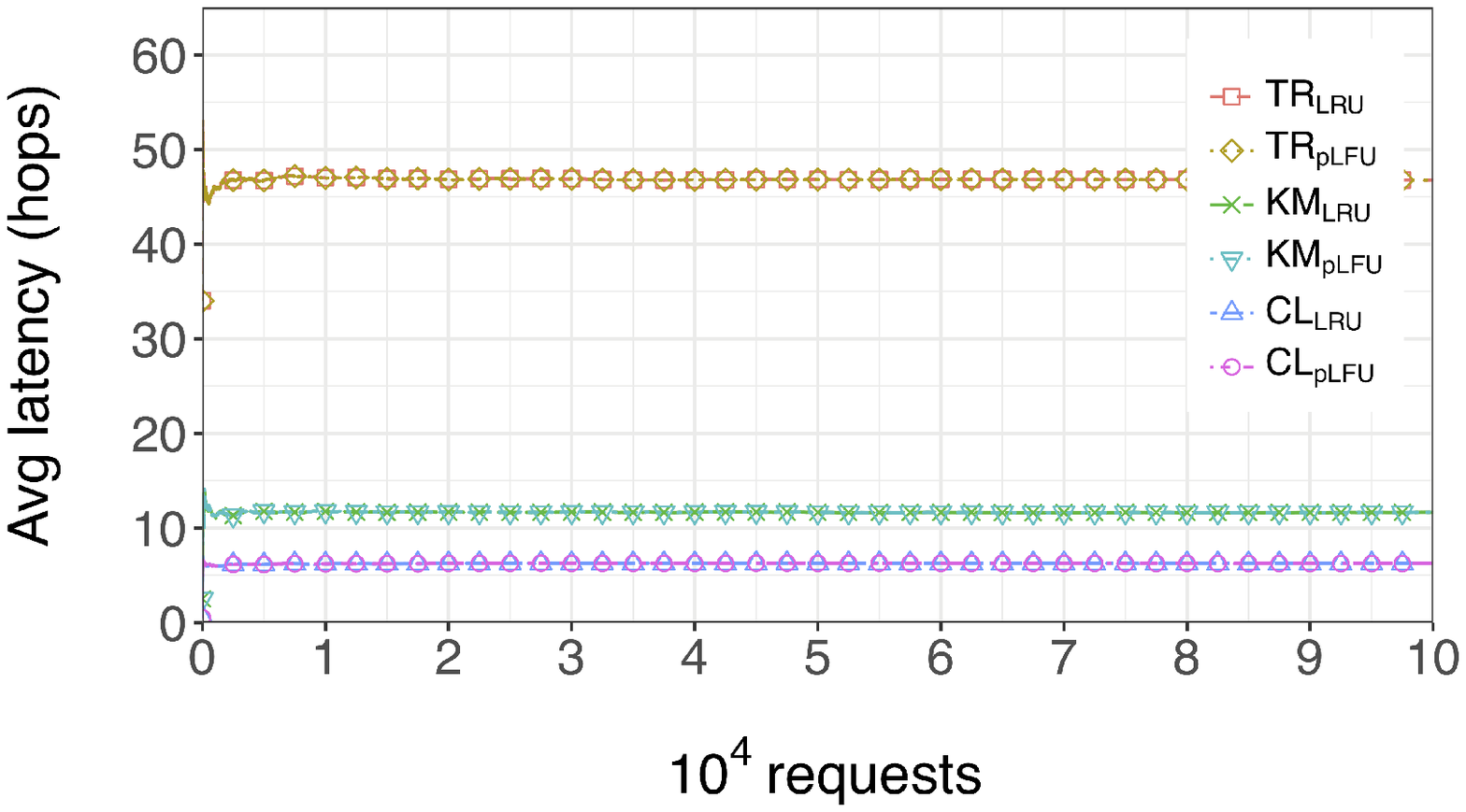}
                \caption{Latency versus homogeneous request intensity}
                \label{fig:hom-nc}\end{subfigure}
                \begin{subfigure}[c]{.4\textwidth}
                \includegraphics[width=\linewidth]{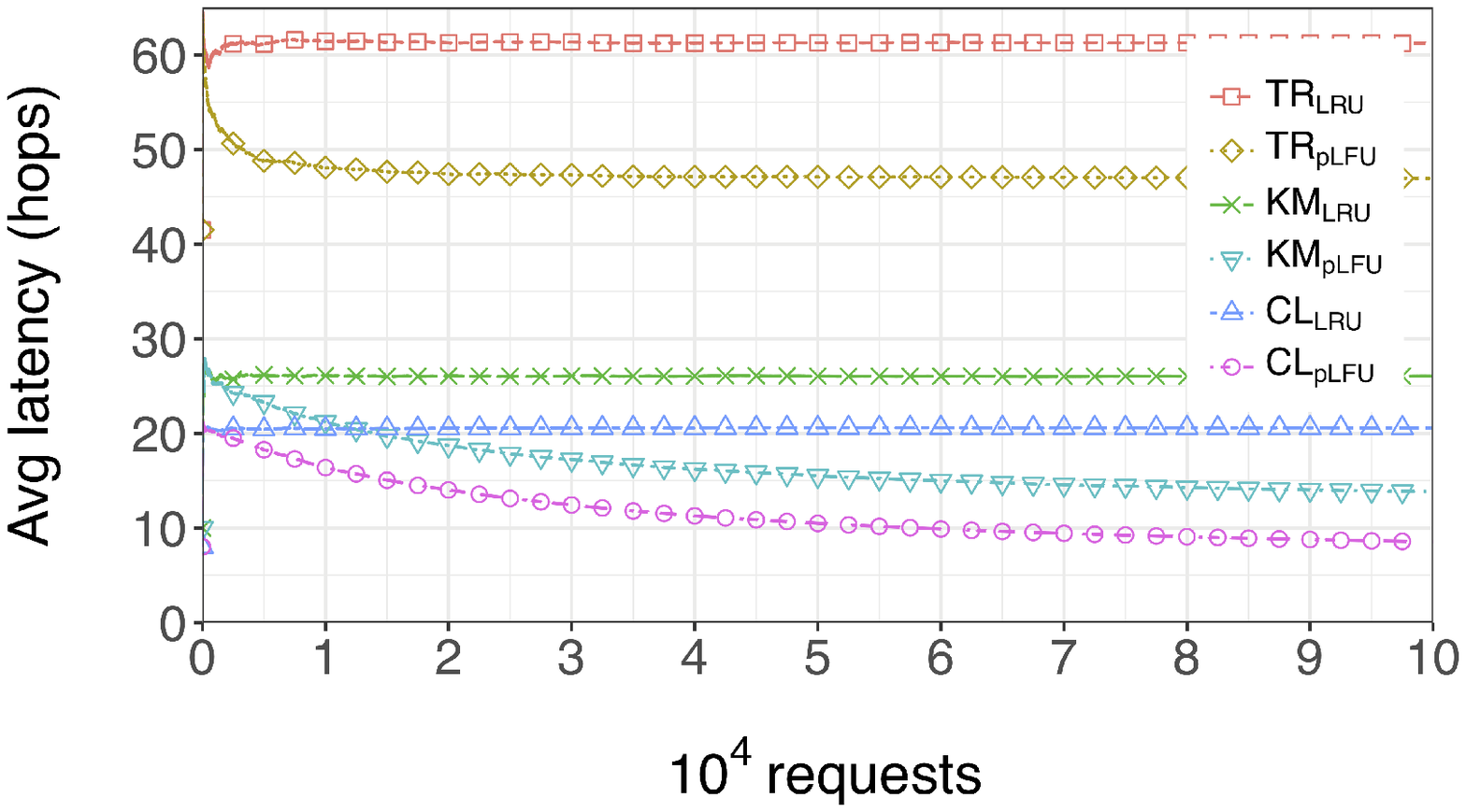}
                \caption{Latency versus heterogeneous request intensity}
                \label{fig:het-nc}
        \end{subfigure}
        \begin{subfigure}[c]{.4\textwidth}
                \includegraphics[width=\linewidth]{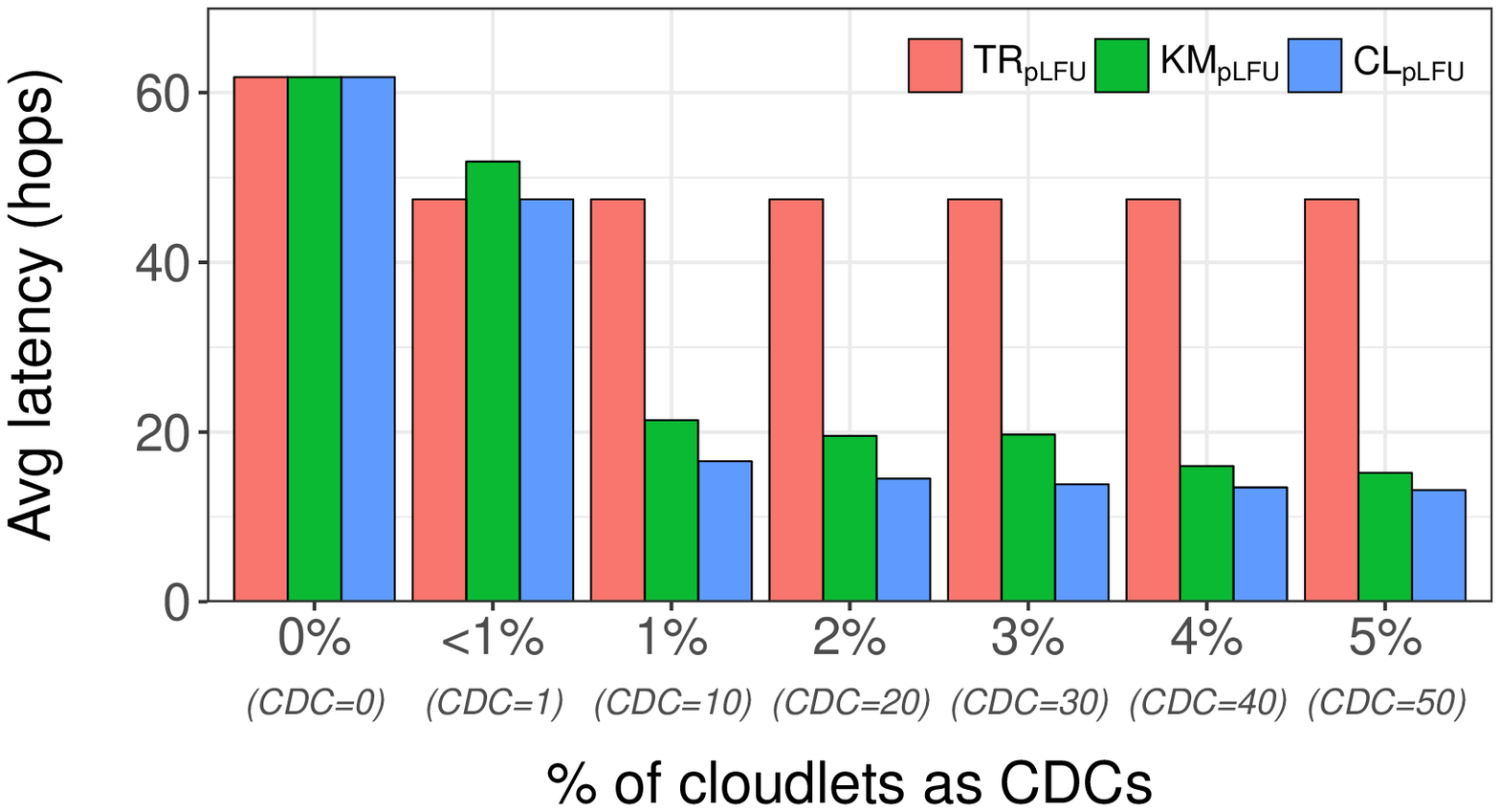}
                \caption{Latency as a function of CDC percentage}
                \label{fig:cdc-percent}
        \end{subfigure}
        \begin{subfigure}[c]{.4\textwidth}
                \includegraphics[width=\linewidth]{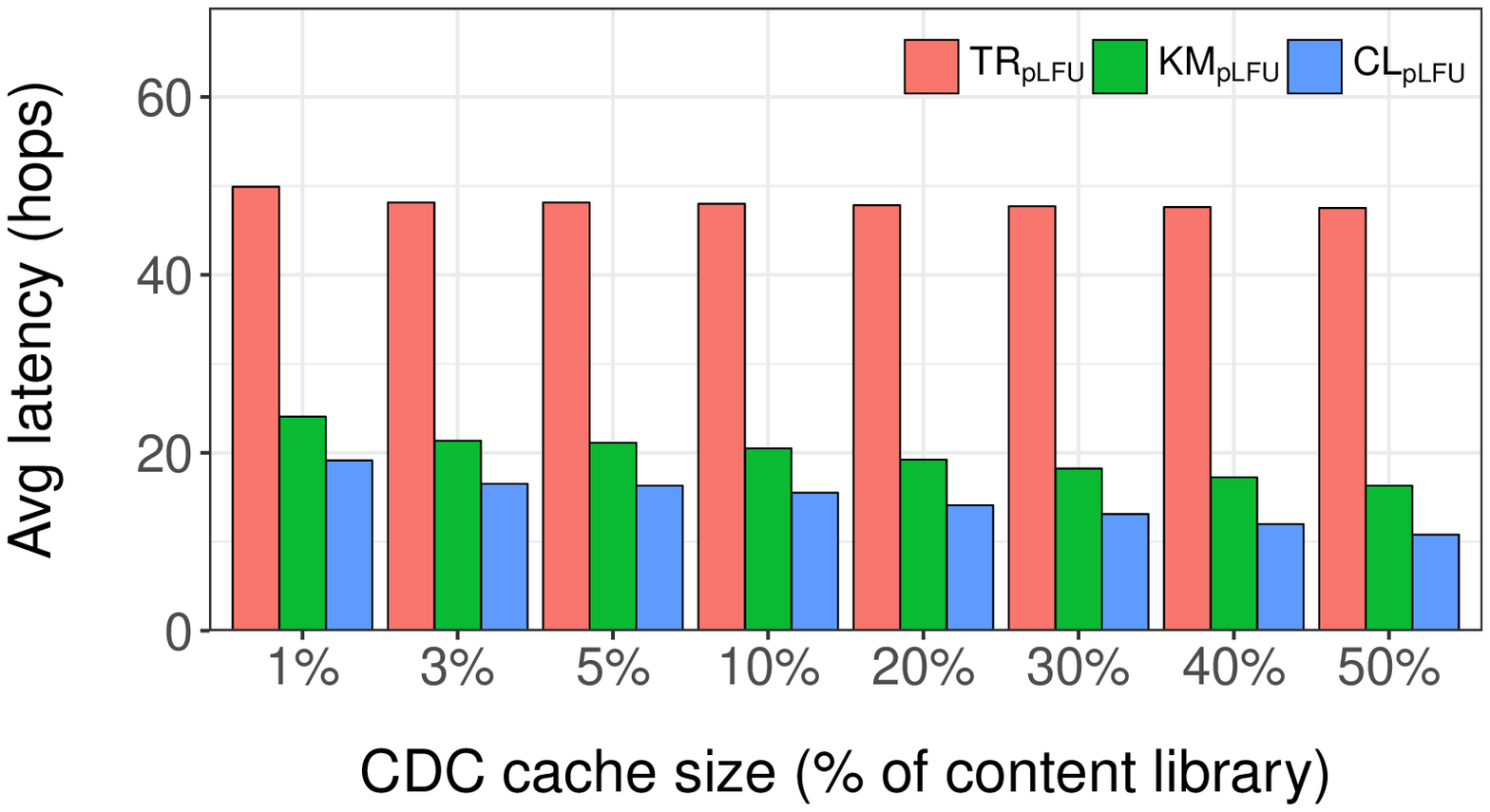}
                \caption{Latency as a function of CDC capacity}
                \label{fig:cache-percent}
        \end{subfigure}
        \caption{\textbf{(a)} and \textbf{(b)} Content-centric LinkNYC performance results comparing LRU to \PLFU~considering homogeneous and heterogeneous content request intensity and using different clustering techniques. \textbf{(c)} Average latency as the number of CDCs increases. \textbf{(d)} Average latency as the capacity of CDCs increases.}
        \label{fig:linknyc-results}
\end{figure*}

\comment{
\begin{table}
\centering
\setlength\tabcolsep{4pt}
\begin{minipage}{0.4\textwidth}
\centering
\setlength{\extrarowheight}{0.2cm}
\begin{tabular}{|x{1cm}|x{0.75cm}|x{0.75cm}|x{0.75cm}|x{0.75cm}|}\hline
\diag{.1em}{1cm}{$f$}{$cdc$}& $cdc_1$& $cdc_2$& \ldots& $cdc_N$\\ \hline
$f_1$&1&0&\ldots&0\\ 
$f_2$&0&0&\ldots&1\\ 
\sbox0{\dots}\makebox[\wd0]{\vdots} &\sbox0{\dots}\makebox[\wd0]{\vdots} &\sbox0{\dots}\makebox[\wd0]{\vdots} &$\ddots$&\sbox0{\dots}\makebox[\wd0]{\vdots} \\ 
$f_M$&1&0&\ldots&1\\ \hline
\end{tabular}
\caption{CDC content store}
\label{tab:cdc-cs}
\end{minipage}%
\hfill
\begin{minipage}{0.4\textwidth}
\centering
\setlength{\extrarowheight}{0.2cm}
\begin{tabular}{|x{1cm}|x{0.75cm}|x{0.75cm}|x{0.75cm}|x{0.75cm}|}\hline
\diag{.1em}{1cm}{$f$}{$cdc$}&$cdc_1$&$cdc_2$&\ldots&$c_N$ \\ \hline
$f_1$&23&10&\ldots&15\\ 
$f_2$&6&17&\ldots&19\\ 
\sbox0{\dots}\makebox[\wd0]{\vdots}&\sbox0{\dots}\makebox[\wd0]{\vdots}&\sbox0{\dots}\makebox[\wd0]{\vdots}&$\ddots$&\sbox0{\dots}\makebox[\wd0]{\vdots} \\ 
$f_M$&37&12&\ldots&43\\ \hline
\end{tabular}
\caption{Window statistics during period $\kappa$}
\label{tab:cdc-ws}
\end{minipage}
\end{table}
}

Our proposed popularity-based content caching method, termed \PLFU, works as follows.
Each cloudlet computes and maintains an estimate of the average number, $\bar{c}_{f}^{(\kappa)}$, of each content $f$'s requests encountered by the cloudlet at the $\kappa^{th}$ period. We propose to compute $\bar{c}_{f}^{(\kappa)}$ as a weighted moving average $\bar{c}_{f}^{(\kappa)} = \alpha \bar{c}_{f}^{(\kappa-1)} + (1-\alpha)c_{f,w}^{(\kappa)}$ where $c_{f,w}^{(\kappa)}$ is the number of requests for content $f$ encountered at the $\kappa^{th}$ period during window $w$, and $\alpha$ is a weighting design parameter set between $0$ and $1$. The parameter $\kappa$ refers to the current index of the content observation window $w$. In practical scenarios content popularity mainly depends on current social trends and time of day. In our work, however, content popularity is assumed to follow the Zipf distribution. Zipf parameters are periodically altered to help mimic changing popularities over time. CDC caching policies in large urban communication networks must be able to quickly adapt to these changes. For this reason a small observation window $w$ (i.e., 10) is needed to quickly detect changes in content popularity over time.
The {\em popularity index} of content $f$ is then computed as $p_{f}^{(\kappa)} = \frac{\bar{c}_{f}^{(\kappa)}}{\sum_{g\in \contentset}\bar{c}_{g}^{(\kappa)}}$ where $\contentset$ is the set of contents the cloudlet has encountered. Subject to cache sizes, CDCs store encountered content with the highest popularity indexes.

Figure~\ref{fig:linknyc-results} illustrates the benefit of a content-centric shift within Brooklyn's LinkNYC network. For this preliminary analysis we leverage Brooklyn's 25 CDCs, from Table~\ref{tab:nyc-summary}, where we assume CDCs are capable of storing up to 3\% of the network's total popular content. We also assume the latency required to fetch content from the original content publisher is minimal. Figures~\ref{fig:hom-nc} and ~\ref{fig:het-nc} depict the average latency experienced of LRU and \PLFU~caching policies while considering content homogeneity and heterogeneity respectively. Although content homogeneity, where all users request the \textbf{same} content through their respective CDCs, is not a realistic case, it does highlight the performance benefit of our population based clustering and CDC placement heuristic from Section~\ref{sec:p6-clustering}, \textit{CL}, compared to traditional, \textit{TR}, and k-means, \textit{KM}, approaches. Since content requests are the same across CDCs, \PLFU~does not provide any added benefit over LRU. However, when more realistic heterogeneous content requests are introduced we see the performance boost of a \PLFU~approach. A traditional single regional CDC employing \PLFU~(TR\textsubscript{\PLFU}) provides a near 25\% reduction in average latency per request as opposed to TR\textsubscript{LRU}.
Introducing in-network caching to LinkNYC provides great improvements to a traditional infrastructure, however, the average latency remains high (47.5 hops) and is not ideal in dynamically changing environments resulting in sub-par user QoE. Incorporating multiple CDCs within LinkNYC addresses this issue by pushing content closer to the edge. Both KM\textsubscript{LRU} and CL\textsubscript{LRU} provide immediate latency reductions of 58\% and 67\% compared to TR\textsubscript{LRU}. Although CDCs have limited capacities, content popularity can be learned in anticipation of future requests. Over time, and as content requests intensify, KM\textsubscript{\PLFU} and CL\textsubscript{\PLFU}
exhibit promising results yielding a 65\% and 85\% latency reduction compared to TR\textsubscript{\PLFU}.
Using the \PLFU~policy, Figure~\ref{fig:cdc-percent} illustrates average latency as the number of CDCs deployed within Brooklyn's topology is increased and a content request intensity of $20\times10^3$. Naturally, latency improves as more CDCs are deployed, however, the improvement is minimal as the percentage of CDCs reaches around 5\%. The effect of increasing CDC cache sizes is shown in Figure~\ref{fig:cache-percent}. Results show that increasing cache sizes alone may not necessarily improve latency since contents still need to traverse to end users. We promote that there exists a balance between the number of CDCs and a CDC's cache size.

In Sections~\ref{sec:p6-cdc} and~\ref{sec:p6-plfu}, a cloudlet-based content placement architecture coupled with a popularity-driven content caching scheme that stores content based on local CDC popularity indexes was proposed. It is shown that integrating content popularity when making content placement and caching decisions reduces both the average downloading time and the network backhaul traffic as opposed to pure reactive policies. Although the popularity-driven caching scheme, \PLFU, yields promising results, it does not guarantee content placement that minimizes the global average latency, as it does not account for content availability and popularity in neighboring CDCs. An amalgam of user mobility, content heterogeneity, and in-network storage introduces several complex challenges. More advanced content-centric solutions must be sought in order to reduce network issues and address underlying limitations such as rising content heterogeneity, user mobility, backhaul load, network failure and deployment costs. Next, we propose a potential solution for overcoming these challenges. Specifically, we propose to exploit and rely on cooperation and information sharing among neighboring CDCs for providing faster content access and lesser backhaul traffic.

\subsection{Cloudlet Cooperation for Faster Content Access}
\label{sec:proactive}

Content caching and placement decisions should depend not only on local but also neighboring CDC conditions and observations, such as content popularity, storage capacity, content availability in the neighborhood, user population, and link/network condition (congestion, data rates, etc.).
Intuitively, when a new content is requested within some local CDC, the decisions on $(i)$ whether to cache the new content or not, $(ii)$ which CDC to cache the content at, and $(iii)$ which existing cache content to evict should all be community-based in that both local and neighboring CDCs should all cooperate and be involved in making such decisions so that globally optimal placements that satisfy users' QoE while accounting for resource availability constraints can be made.
For example, if the new content is available at a nearby CDC, then there might not be a need for caching it again at the local CDC, thus saving local cache resources yet while still allowing users to receive the content with acceptable latency.
Now if the new content is not available locally, nor in neighboring CDCs, then the decision to whether to cache or not should depend on its community popularity, not just its local popularity.
If this content is popular enough to cache, then the decision to where it should be cached should weigh in its popularity indexes at the different CDCs within the network. That is, even if the content is being requested by a user located within a CDC $i$, it might be more efficient to cache it at a neighboring CDC if popularity indexes show that the content is more likely to be requested within the neighboring CDC and/or the neighboring CDC has more available cache space.

Designing cooperative content caching and placement approaches that consider the aforementioned performance aspects has not been fully addressed by the research community. And deriving models that capture the various content and network aspects influencing these decisions, such as content popularity, storage availability, content availability, user population, and network condition, is a challenging task that requires careful study. In this work, we propose to introduce a score function $\util_{f,i}^{(\kappa)}$ that each CDC $i$ maintains for each of its encountered content $f$, updated every period $\kappa$, and used to make content placement and caching decisions. Here, $\util_{f,i}^{(\kappa)}$ represents the cost associated with caching content $f$ at CDC $i$ at the update period $\kappa$. We propose that this function captures and models the following aspects:

\begin{itemize}
 \item \textbf{Content popularity ($p_{f,i}^{(\kappa)}$):} Popularity index of content $f$ as observed by CDC $i$ during update window $\kappa$.

 \item \textbf{Content availability ($a_{f,i}^{(\kappa)}$):} An availability binary index of content $f$, where 1 indicates that $f$ is cached in CDC $i$ during update window $\kappa$, and 0 otherwise.

 \item \textbf{Population density ($r_{i}$):} This reflects the population density of CDC $i$, as described in Section~\ref{sec:p6-pop}.

 \item \textbf{Inter-CDC delay ($\ell_{i,j}$):} It represents the delay experienced by a user belonging to a CDC $i$ requesting content cached at a neighboring CDC $j$. It essentially captures the number of hops, as well as the link bandwidth capacity of each hop, connecting CDCs $i$ and $j$.

 \item \textbf{Intra-CDC delay ($\ell_{i}$):} It represents the average delay experienced by a user requesting content from its community CDC $i$.

\end{itemize}

We propose to model $\util_{f,i}^{(\kappa)}$ as a weighted average of a neighborhood score and a local score as such:
\begin{equation*}
\util_{f,i}^{(\kappa)}=\beta\times\network_{f,i}^{(\kappa)} + (1-\beta)\times\node_{f,i}^{(\kappa)}
\end{equation*}
where $\network_{f,i}^{(\kappa)}=\frac{\sum_{j \in \Neigh_i}{r_j p_{f,j}^{(\kappa)}\ell_{i,j}^{(\kappa)}}}{|\Neigh_i|}
\mbox{\;\;and \;\;}
\node_{f,i}^{(\kappa)}=\frac{r_i p_{f,i}^{(\kappa)}}{\ell_{i^{(\kappa)}}}$
where $\Neigh_i$ is the set of CDC $i$'s neighboring CDCs and $\beta$ is a design parameter set between 0 and 1. $\util_{f,i}^{(\kappa)}$ captures the local and neighborhood benefit, from the perspective of the deciding CDC $i$, of caching content $f$. That is, as content traverses the network each CDC decides to store the traversing content based on its score value. Considering local and neighborhood characteristics allows for our score function to better service dynamic content environments. Deciding on the number of neighboring CDCs to consider is a design choice discussed in Section~\ref{sec:p6-analysis}.

Note that additional score functions can be modeled to capture CDC resources (processing, storage, memory, energy, etc.) more accurately and are left for future investigation. The neighborhood score function, $\network_{f,i}^{(\kappa)}$, essentially represents a weighted average latency that user requests generated within CDC $i$ will experience if content $f$ is cached among its neighboring CDCs. $\node_{f,i}^{(\kappa)}$ represents an average latency if content $f$ is stored locally in CDC $i$. For the extreme case when $\beta=1$, content resulting in higher potential latencies among our neighbors will be locally stored more often. On the other hand, when $\beta=0$, our score function stores content with the highest local popularities and low latencies. That is, highly popular content with low latencies to fetch are favored over others.

\comment{
To assist in the score formulation, we propose the construction of virtual CDC content layers. Assuming boroughs contain a subset of popular Internet content, virtual layers are formed indicating where each content is stored as shown in Figure~\ref{fig:virtual}. Virtual layers are constructed based on the CSs maintained and periodically shared by CDCs with remaining cloudlets as content traverses the network. Naturally if content is not available within a respective community, it can be requested through virtual CDCs. Otherwise, content is requested from the original publisher. CDCs can use the resulting virtual layers to sift through cloudlets and efficiently compute the neighborhood portion of the score.

\begin{figure}
 \centering
 \includegraphics[width=.375\textwidth]{virtual_layers_3.eps}
 \caption{\label{fig:virtual}Virtual CDC content depiction}
\end{figure}
}

\subsection{Content Popularity Skewness}

Naturally, content popularity within communities can vary over time and in order to account for this dynamicity, tuning the design parameter $\beta$, in $\util_{f,i}^{(\kappa)}$, becomes especially useful and can provide a more responsive storage mechanism. We assume content popularity follows the Zipf distribution, $f(\tau,s,M)=\frac{1/\tau^s}{\sum_{m=1}^M(1/m^s)}$, where $\tau$ is a content's rank in terms of popularity, $M$ is the total number of contents, and $s$ controls the skewness of the distribution. The Zipf distribution is widely used in the literature to describe popularity~\cite{Sarrar:zipf:2012,Faloutsos:zipf:1999,Adamic:zipf:2004}.

\begin{figure}
 \centering
 \includegraphics[width=.4\textwidth]{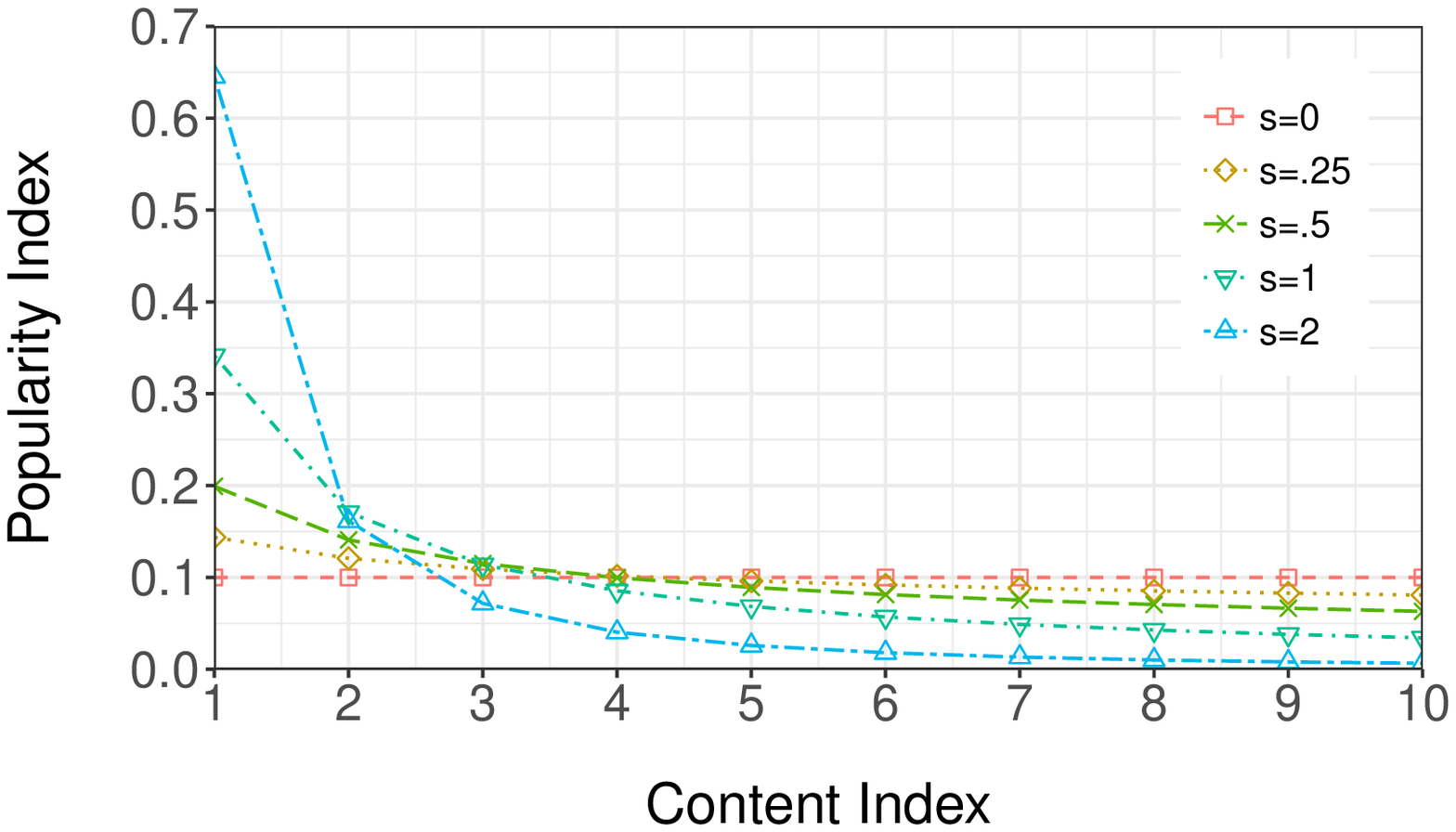}
 \caption{\label{fig:s_pop}Popularity indexes as a function of $s$}
\end{figure}

It is clear from Figure~\ref{fig:s_pop} that as parameter $s$ increases so does the skewness in content popularity especially when $s\geq1$. We propose a dynamic parameter $\beta$, in the score function $\util_{f,i}^{(\kappa)}$, that is inversely proportional to parameter $s$. That is, for the extreme case when contents have equal popularities (i.e., $s=0$) simply storing the most popular content locally, as done with~\PLFU, will not suffice. However, querying content popularities and availabilities among neighboring CDCs for storage decisions intuitively improves responsiveness by effectively consolidating storage capabilities. Thus, a higher $\beta$ value is preferred to favor our neighborhood in our cache decisions. Conversely, when content skewness is high (i.e., $s=2$), it is more efficient to locally store the most popular content within a community. In other words, as content popularities are more skewed our score function favors storing the most popular content locally. For this reason a dynamic $\beta$ parameter is necessary to balance between local and neighborhood based storage decisions depending on the $s$ parameter in the Zipf distribution. This allows $\util_{f,i}^{(\kappa)}$ to be adjusted accordingly and adapt to network changes over time.

\subsection{Cooperative Content Placement}

The parameter $\beta$ allows us to balance between the need for having responsive content delivery locally versus among our CDC neighbors. This is especially useful for dynamic content popularity within large urban communication networks. As these above network and content conditions change over time, each CDC must periodically maintain and compute score function values for encountered contents. This could be done by having CDCs query neighboring CDCs for content popularity indexes, population densities, CDC latency and resource availability, and use this information for updating these values, which are then used as follows for content placement decisions:

\begin{itemize}

\item Content popularity indexes, $p_{f,i}^{(\kappa)}$, are maintained for each encountered content $f$ during period $\kappa$ as described in Section~\ref{sec:p6-plfu}

\item For each content $f$ encountered at CDC $i$, the $\network_{f,i}^{(\kappa)}$ and $\node_{f,i}^{(\kappa)}$ portions of the content score value, $\util_{f,i}^{(\kappa)}$, are computed and maintained during period $k$

\item For each CDC $j$ in our set of neighbors $\in\Neigh$, inter-CDC latency, $\ell_{i,j}^{(\kappa)}$, is computed based on content $f$'s availability among our neighbors. Note, the additional latency required to fetch content from the original publisher is also captured:

\begin{equation*}
    \ell_{i,j}^{(\kappa)} = \begin{cases}
               \ell_{i,j}+\ell_{j,origin}^{(\kappa)}   & a_{f,i}^{(\kappa)} = 0\\
               \ell_{i,j}               & a_{f,i}^{(\kappa)} = 1\\
           \end{cases}
\end{equation*}

\item Intra-CDC latency, $\ell_{i}^{(\kappa)}$, is computed during period $\kappa$ as the average latency (i.e, hops) from each cloudlet to its community CDC $i$. Similarly, in the event that $a_{f,i}^{(\kappa)} = 0$, $\ell_{j,origin}^{(\kappa)}$ is accounted for to capture the additional latency required to fetch content from the original publisher

\item Design parameter $\beta$ is set based on an estimation of the $s$ during period $\kappa$ and maintained per CDC

\item For each content $f$ encountered at CDC $i$, it is stored locally if $f\in\util_{max}$ where $\util_{max}$ is the set of contents with the highest local content scores, subject to CDC storage capacity. Otherwise no caching takes place. If CDC $i$'s storage is full and $f\in\util_{max}$ then the content with the minimum score value is evicted from the local cache

\end{itemize}

Considering both local and neighborhood conditions and content availability in modeling $\util_{f,i}^{(\kappa)}$ allows us to account for changing content popularities within different communities over time. Our intuition indicates that storage decisions must rely heavily on a CDC's content popularity distribution. That is, as content popularity becomes more uniform, considering content availability among our neighbors improves performance. However, as content popularity is more skewed, storage decisions should be based on local CDC content popularities. We discuss our design choices and analyze our results next.

\section{Framework Analysis: A LinkNYC Use-Case}
\label{sec:p6-analysis}

A detailed examination using LinkNYC's urban communications network is performed to investigate the fundamental benefits of shifting from a traditional IP network to a content-centric network where popular content is cached as it disseminates throughout the network. \texttt{ndnSIM} is an NDN simulator based on NS-3 and was used to analyze the proposed urban communications network infrastructure. Traditionally, content delivery nodes are distributed globally to bring content as close to the geographic location of the user. However, promising results show that having multiple content-delivery cloudlets within LinkNYC's infrastructure coupled with smarter caching techniques dramatically improves overall network performance and stability. In an increasingly mobile and dynamic environment, where network conditions and service continuity issues are amplified due to user mobility and varying content interests; network stability, efficiency and content delivery responsiveness are essential to a user's QoE.

\subsection{Setup}

In this analysis we focus on NYC's Brooklyn borough. Among the topology's 1004 cloudlets, 25 CDCs are geographically selected based on population densities using our technique from Section~\ref{sec:p6-rtop}. Content requests are then generated following a Zipf distribution with varying $s$ parameters per community to realistically simulate community-specific interests and changing content popularity. We assume LinkNYC's popular content library, $M$, contains 600 contents categorized by type (i.e., Sports, Entertainment, Politics and Education) where each CDC's storage capacity, $C_i$, is set to 20 contents. Content sizes are assumed to be constant where popularity indexes follow the Zipf distribution. However, fluctuating population densities and varying link capacities are currently in progress and have been left for future work. Requesting content from its original publisher also yields varying latencies per community (i.e., between 250 and 500 hops). In addition, popularity skewness and interests are periodically shuffled among communities to capture time variation within the LinkNYC infrastructure. Finally, design parameters $\alpha$ and $\beta$ are chosen to control (i) the framework's sensitivity to popularity changes and (ii) the degree to which we balance between neighborhood and local content storage decisions respectively. Our simulation parameters are summarized in Table~\ref{tab:simparam}.

\begin{table}[]
\begin{center}
\begin{tabular}{ll}
\toprule
Parameter & Value \\
\midrule
	Borough & Brooklyn \\
    Cloudlets & 1004 \\
    CDCs & 25 \\
    Content Categories & Sports, \\
    & Education, \\
    & Politics, \\
    & Movies \\
    $|M|$ &  600 \\
    $C_i$ & 20 \\
    Number of requests & $10^6$ \\
	$\ell_{origin}$ & $[250,500]$ hops    \\
    $|\Neigh_i|$ & 24 \\
    $s$ range & $[0,2]$ \\
    $\hat{s}$ observations & $10^3$ \\
    $w$ & $10^2$ \\
    $\alpha$ & .2 \\
    $\beta$ &  $1-\frac{1}{1+e^-20\left(s-.5\right)}$ \\
	\bottomrule
\end{tabular}
\caption{Simulation parameters}
\label{tab:simparam}
\end{center}
\end{table}

\subsection{CDC Neighborhood Size}

\begin{figure}
        \centering
        \begin{subfigure}[c]{.4\textwidth}
                \includegraphics[width=\linewidth]{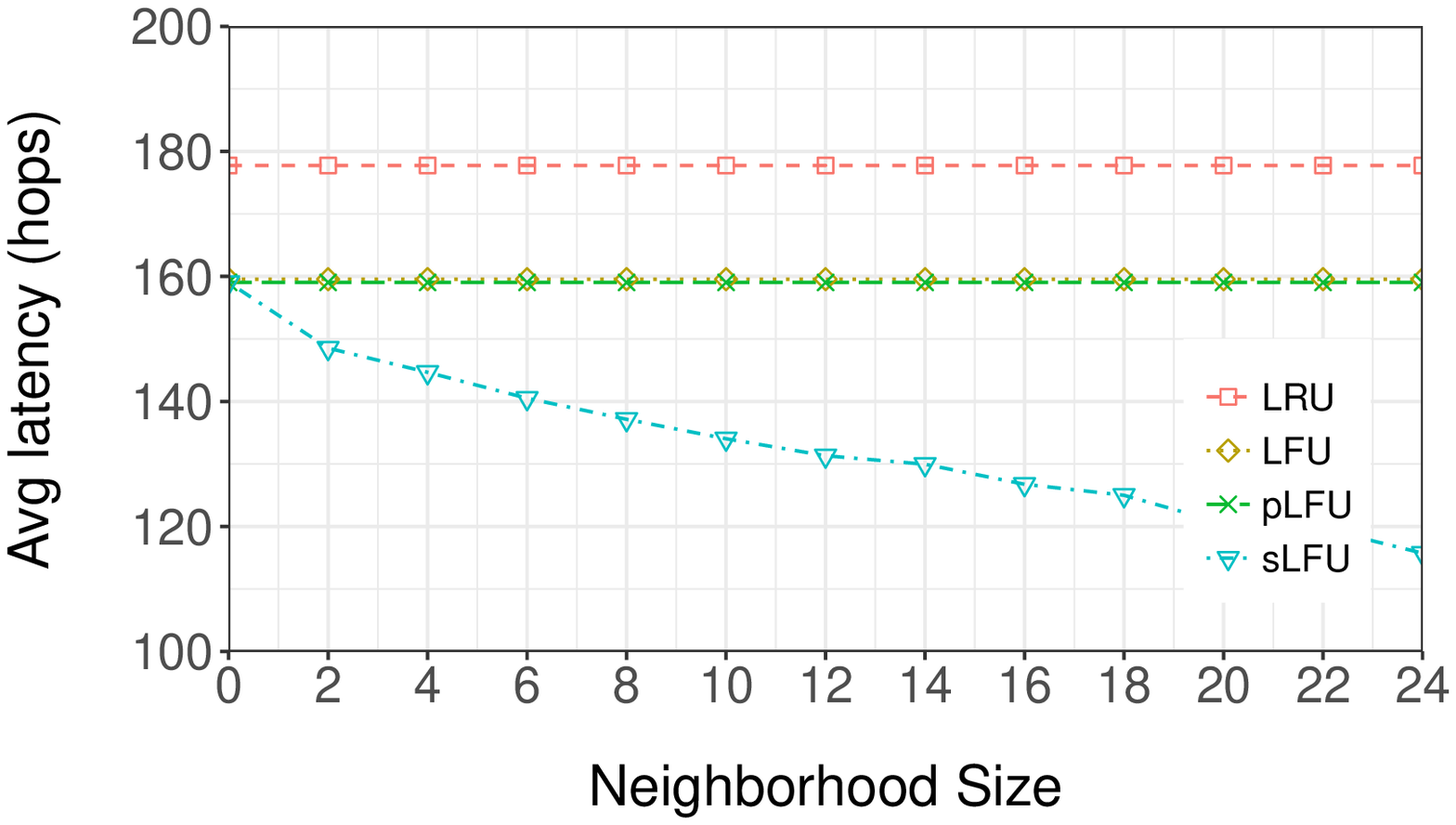}
\caption{\label{fig:neigh}Latency as a function of \ULFU~neighborhood size}
                \label{fig:neigh_lat}\end{subfigure}
                \begin{subfigure}[c]{.4\textwidth}
                \includegraphics[width=\linewidth]{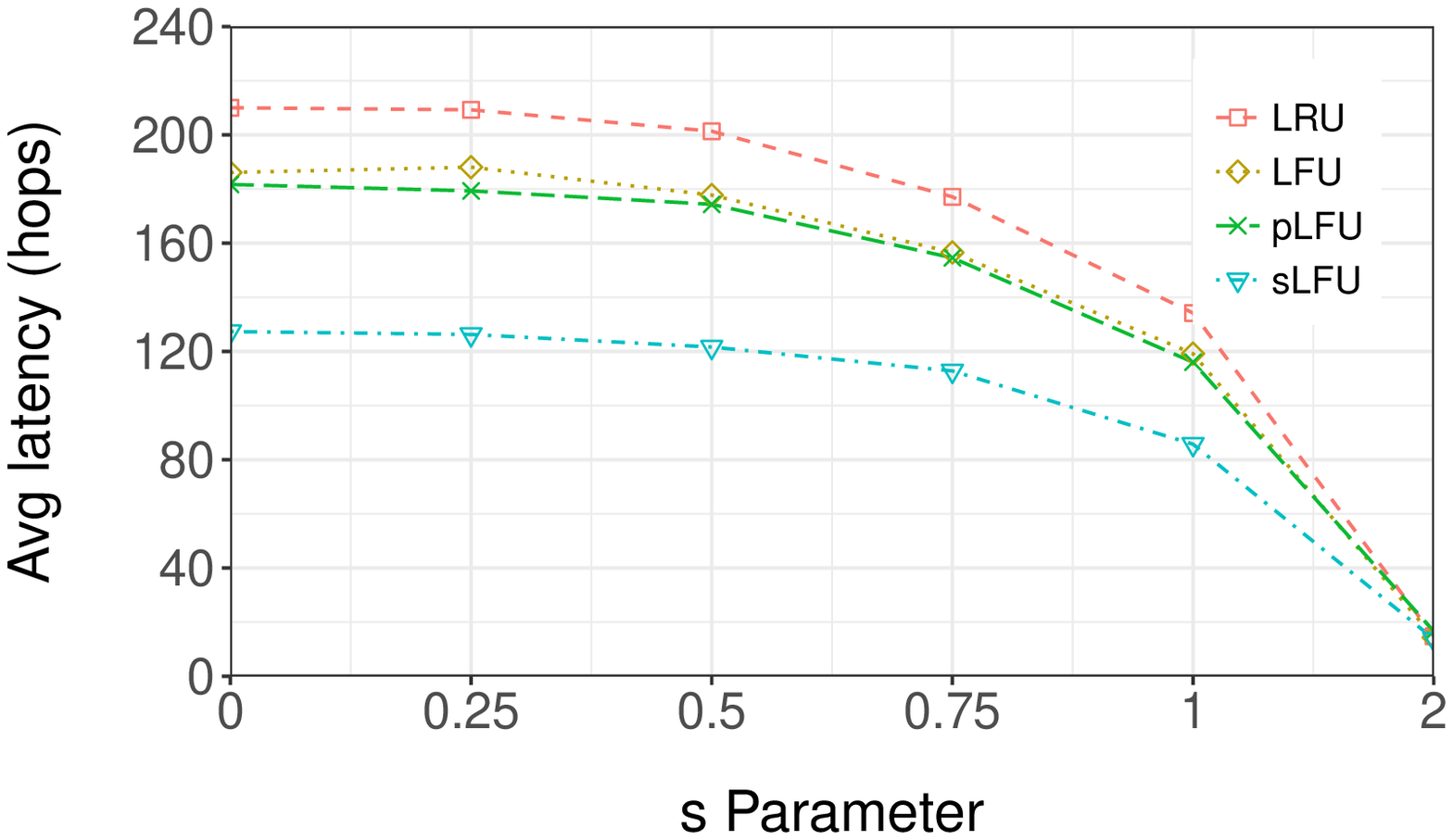}
                \caption{\label{fig:s_pop_res}Latency as a function of different $s$ parameters}
                \label{fig:s_lat}
        \end{subfigure}
        \caption{\textbf{(a)} \ULFU~performance as CDC neighborhood size is increased. \textbf{(b)} Comparing traditional cache replacement policies with \PLFU~and \ULFU~considering static content popularity skewness.}
        \label{fig:neigh_s}
\end{figure}

Neighborhood size can play an important role in overall network performance. Increasing the size of CDC $i$'s neighborhood, $\Neigh_i$, improves performance by effectively consolidating CDC capabilities. Figure~\ref{fig:neigh} emphasizes this performance increase with respect to latency as the neighborhood size is grown to 24. This allows for efficient cooperation among CDCs when making content storage decisions. Since the number of CDCs are relatively low, the negative impact of increasing neighborhood size is often negligible. However, in networks comprising of hundreds of thousands or even millions of devices, such as Massive Internet of Things (MIoT) networks, increasing the neighborhood size can be disadvantageous in terms of complexity due to the added overhead injected into the infrastructure through the exchange of content stores and popularity indexes~\cite{dohler20165g}. LinkNYC network administrators should decide on appropriate neighborhood sizes depending on the application. For instance, during high profile sporting events or tense political climates, many contents may be popular throughout the entire NYC borough and thus using a neighborhood size equal to the total number of CDCs may be preferred. Conversely, geographically restricted interests such as social events may cause only one or two communities to share common interests and thus smaller or even dynamic neighborhood sizes may be beneficial to the overall performance of the network.

\subsection{Impact of Content Popularity}

Figure~\ref{fig:s_pop_res} illustrates the effect content popularity has within LinkNYC. We compare our proposed techniques, \PLFU~and \ULFU, with traditional LRU and LFU cache replacement policies. Here we assign the same $s$ value to all Brooklyn communities to emphasize the importance of tuning the design parameter $\beta$ in $\util_{f,i}^{(\kappa)}$ based on the content distribution. With $s$ values close to zero, $\beta$ is adjusted accordingly allowing for \ULFU~to cooperate more with neighboring CDCs to make storage decisions, drastically reducing latency by up to 33\% compared to LFU and \PLFU, and up to 43\% compared to LRU. However, as we increase $s$, and in turn content popularity skewness,~\ULFU~appropriately relies more on local CDC popularity indexes for storage decisions. That is, when content popularity is very skewed (i.e. $s\geq2$), it is sufficient to only store the most popular content locally; hence, traditional methods also perform well.

\subsection{Content Dynamicity and Estimation}

\begin{figure*}
        \centering
        \begin{subfigure}[c]{.4\textwidth}
                \includegraphics[width=\linewidth]{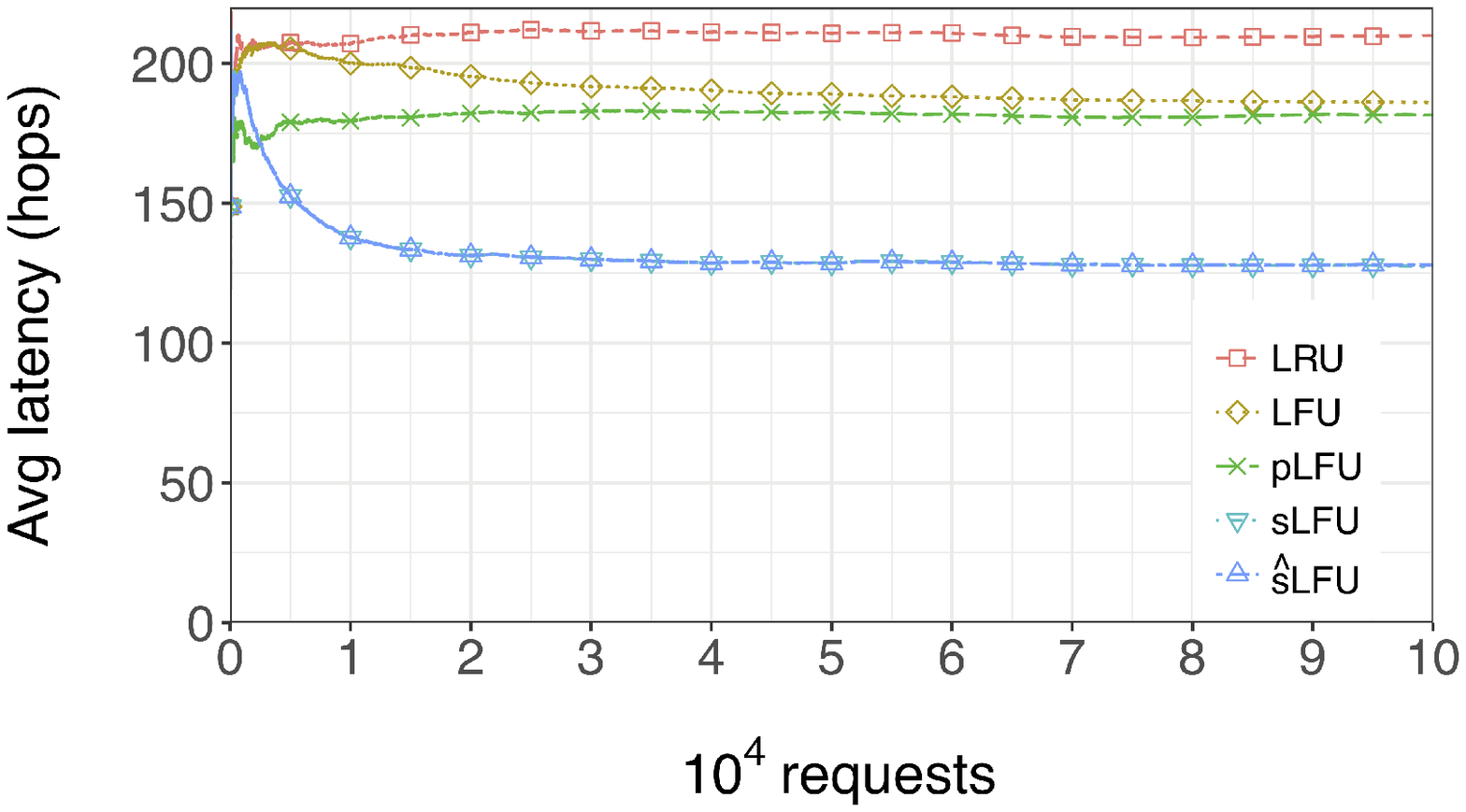}
				\caption{\label{fig:d_s0}Maximum popularity skewness $s=0$}
		\end{subfigure}
        \begin{subfigure}[c]{.4\textwidth}
                \includegraphics[width=\linewidth]{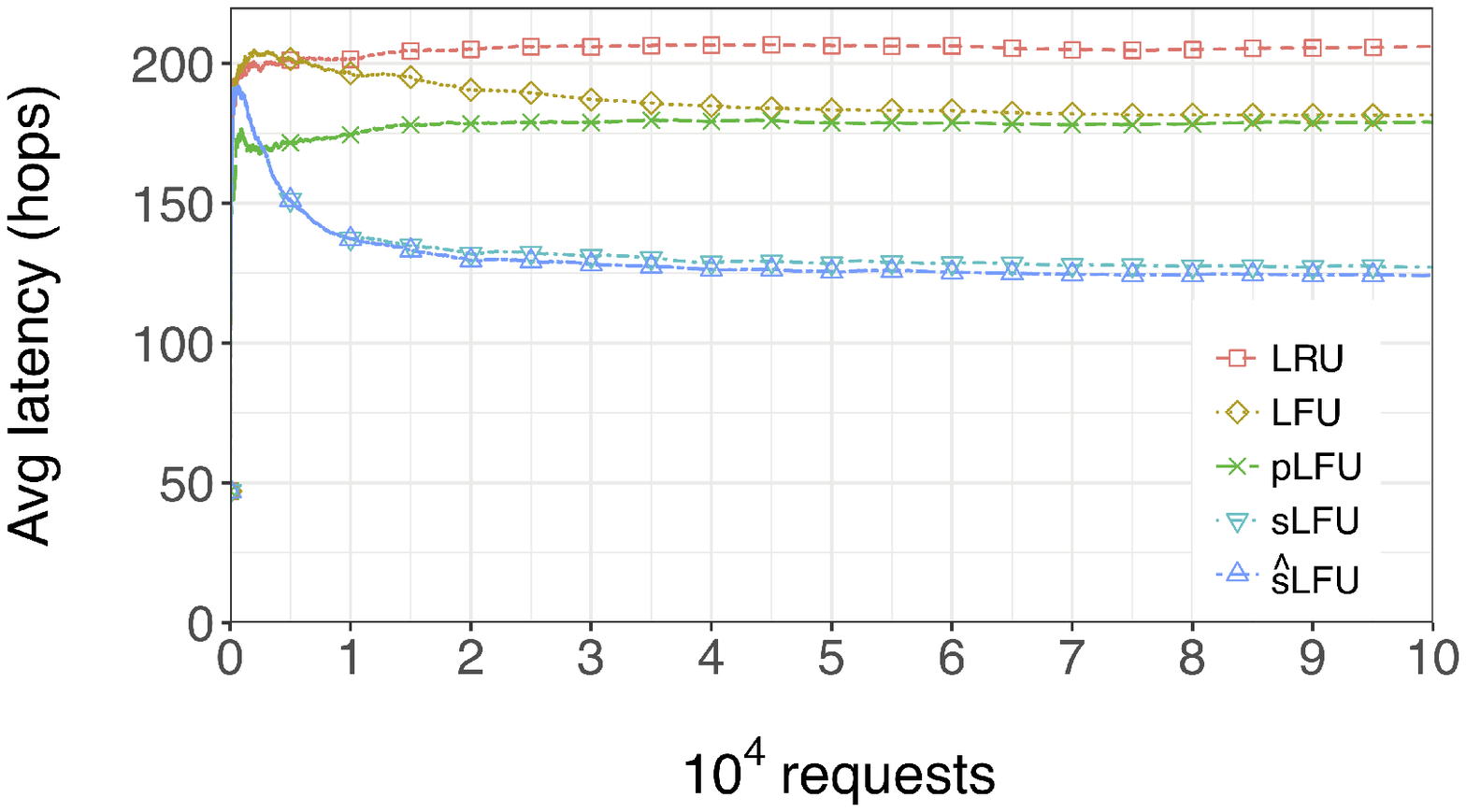}
                \caption{\label{fig:d_s05}Maximum popularity skewness $s=.5$}
        \end{subfigure}
        \begin{subfigure}[c]{.4\textwidth}
                \includegraphics[width=\linewidth]{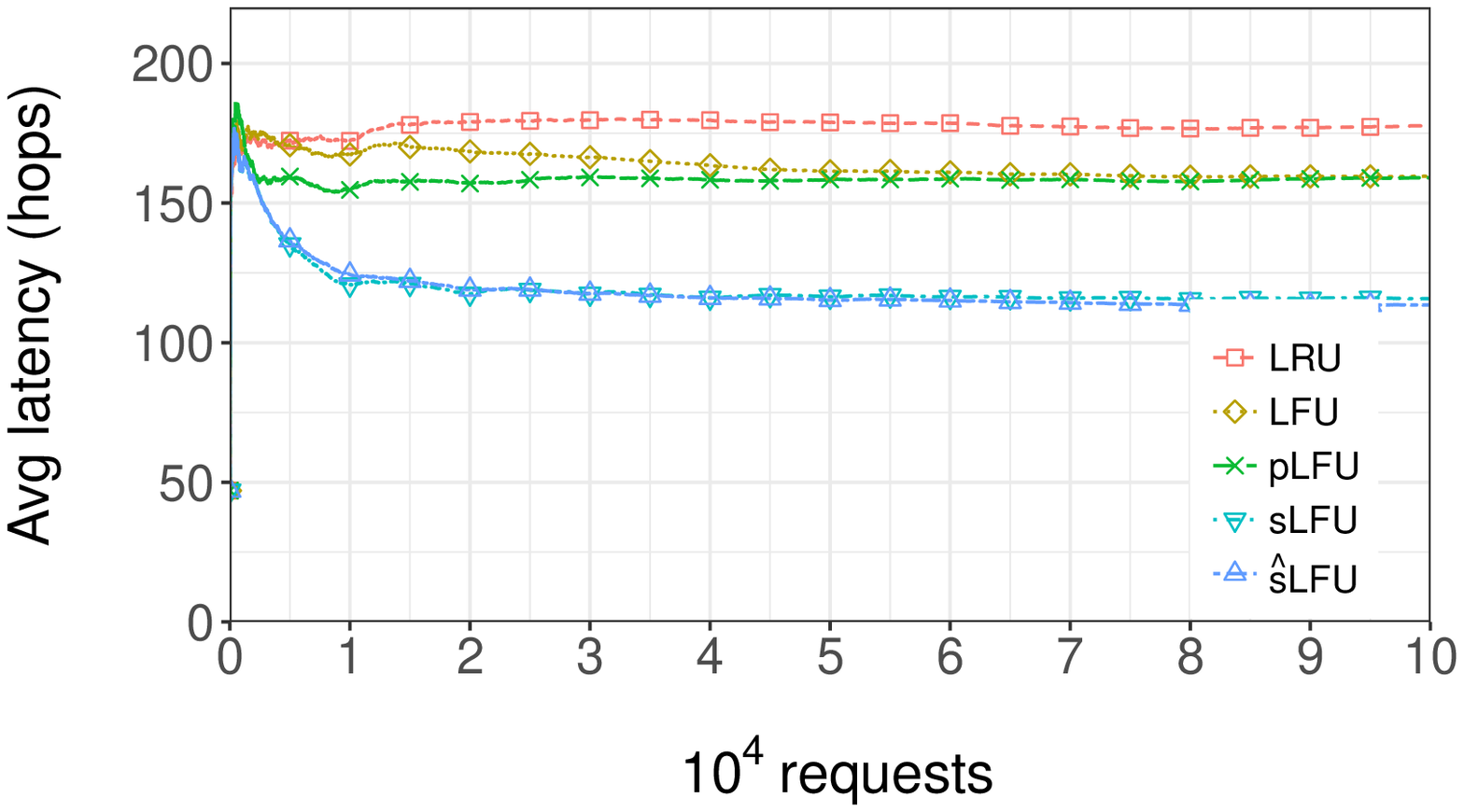}
				\caption{\label{fig:d_01}Maximum popularity skewness $s=2$}					\end{subfigure}
        \begin{subfigure}[c]{.4\textwidth}
                \includegraphics[width=\linewidth]{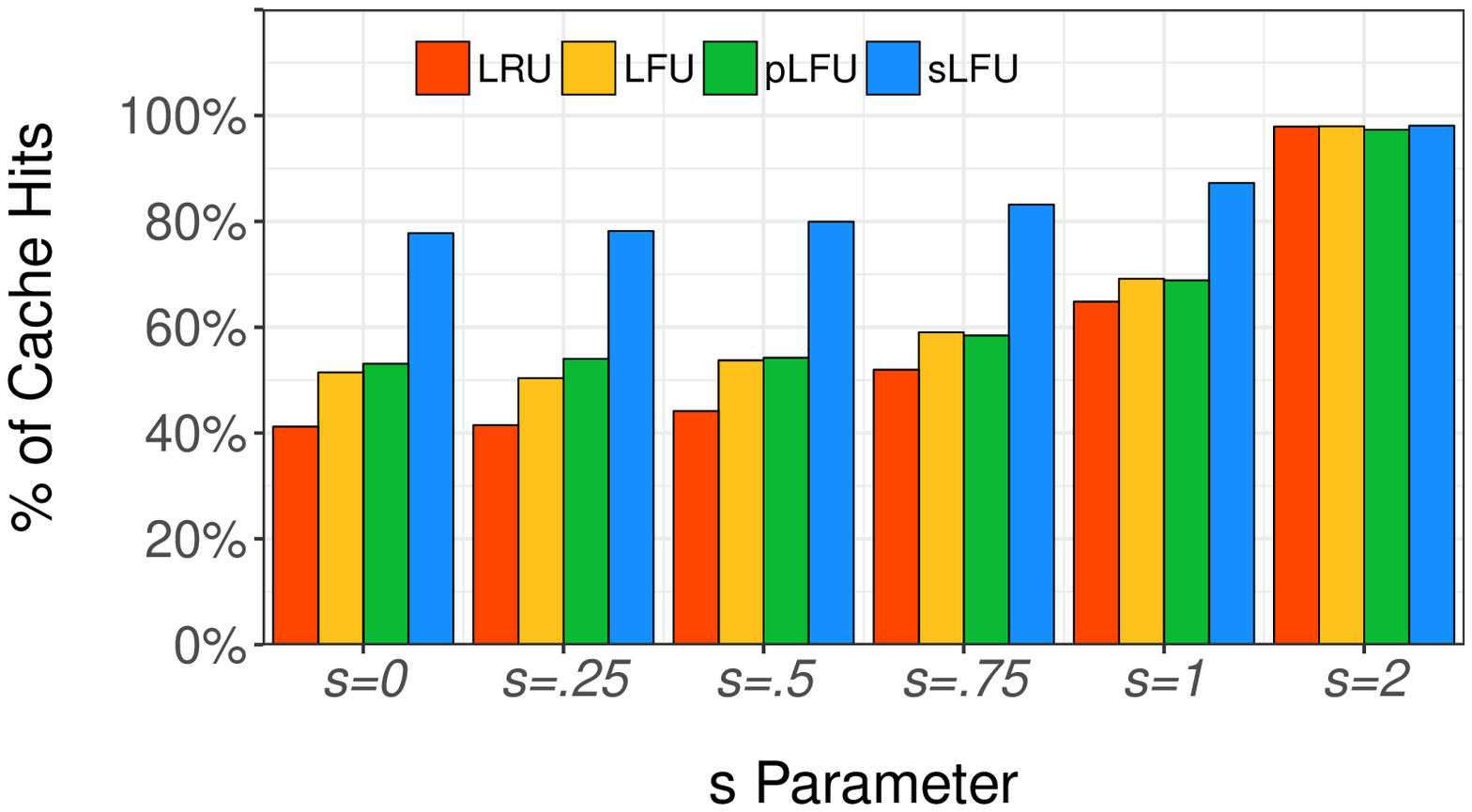}
                \caption{\label{fig:cratio}Ratio of cache hits to misses}
        \end{subfigure}
        \caption{\textbf{(a)},~\textbf{(b)}, and~\textbf{(c)} Illustrate content-centric LinkNYC performance results comparing LRU, LFU, \PLFU~and \ULFU~considering a dynamic network with periodically fluctuating content interests and popularities where $s$ is changed per community with a range between 0 and 2. \textbf{(d)} The ratio of cache hits to cache misses with respect to $s$.}
        \label{fig:s_dynamic}
\end{figure*}

Figure~\ref{fig:s_dynamic} introduces content dynamicity within our content-centric LinkNYC network. Content popularity skewness and interests are periodically shuffled among communities to capture time and content variation. This depicts shifting user interests per community over time (i.e. different sporting events, changing political climates, etc.). During period $\kappa$, where $w$ contents have been encountered, popularity indexes can be leveraged to estimate the $s$ parameter of the content distributions. Each CDC $i$ can then incorporate its estimated $s$ value for tuning design parameter $\beta$ of the score function $\util_{f,i}^{(\kappa)}$. We use maximum likelihood estimation (MLE) to find $s$ based on our current content observations (i.e. $10^3$ observations) allowing \estULFU~to adapt to changing popularities over time.

Figures~\ref{fig:d_s0} and~\ref{fig:d_s05} depict LinkNYC performance as content requests intensify and while popularity skewness is relatively low. The LRU cache replacement policy is the worst performer and results in high average latency, over 200 hops, that is not suitable for a dynamic and mobile environment. LFU and \PLFU~perform similarly, however, \PLFU is able to converge quicker to an average latency of nearly 180 hops. \ULFU's~and \estULFU's cooperation with neighboring CDCs allows them to perform noticeably better achieving over a 30\% improvement compared to traditional methods even in less than ideal conditions where content popularity is relatively similar across communities. When maximum content popularity skewness is increased (i.e. $s=2$), as shown in Figure~\ref{fig:d_01}, traditional methods perform better but are still no match for \ULFU~and \estULFU which still provide a 30\% reduction in average latency. Figure~\ref{fig:cratio} illustrates the cache hit percentages when different $s$ parameters are used. \ULFU~consistently achieves a high cache hit percentage than traditional methods constantly achieving a near 80\% hit rate even with low $s$ values. \ULFU's cooperation-based technique is a novel approach for content placement in large urban communication networks. Implementing \ULFU~at CDCs provides an efficient and cooperative LinkNYC infrastructure for responsive content delivery in dynamically changing environments; essential for mobile users where timely delivery is of the essence.

\subsection{User Mobility}

Mobile users within LinkNYC are bound to experience intermittent connectivity as multiple cloudlets are encountered risking frequent disconnects and disruptions in a user's service especially near the edges of a cloudlet's coverage area~\cite{hassan:15}. This causes potential packet losses, increased response times and re-transmissions. User mobility and dynamic content interests introduce added complexity to an already challenging issue. In order to accommodate mobile users as well as avoid interfering with popular community content we propose the consolidation of CDC storage with a smaller, temporary $L1$ storage strictly used to store prefetched mobile user content on candidate CDCs located on a mobile user's path. Once a mobile user consumes the prefetched content the $L1$ cache is immediately vacated in preparation for other mobile users. In our previous work, we designed a cloudlet-aware mobility based solution to address service continuity issues as multiple cloudlets are encountered across a path.
In~\cite{sinky2016cloudlet} a proactive algorithm is designed and applied by a mobile user as it moves within a community of cloudlets to prefetch content chunks. Prefetching content on candidate CDCs or cloudlets across a path has been shown to address mobility induced service continuity issues.

\section{Conclusion and Future Directions}
\label{sec:p6-conclusion}

In this article, we investigated the fundamental benefits of shifting a large urban communications network to a content-centric network where popular Internet content is cached as it disseminates throughout the network. Different caching techniques were designed and evaluated. Initially, a technique to learn and store the most popular content locally at individual CDCs is proposed. However, results show that performance can be improved through cooperative interaction among neighboring CDCs in order to consolidate storage capabilities. Promising results show that an amalgamation of in-network caching, the placement of multiple CDCs and advanced storage techniques employed within an urban network's infrastructure dramatically improves overall network performance and responsiveness. This type of network unlocks a variety of applications and open challenges. An economical aspect can be introduced where CDC storage space is rented to various businesses to serve popular Internet content or customer services while providing an additional source of revenue for the network. Moreover, LinkNYC can be exploited through mobile cloud computing by augmenting CDC computing capabilities. The focus is still latency minimization while allowing users to offload heavily computational tasks. Our framework lays the foundation for a first-of-its-kind content-centric urban communication network with the focus of minimizing latency and leveraging interest-based content caching through popularity learning and cooperative content placement.
\bibliographystyle{IEEEtran}
\bibliography{main}

\begin{thebibliography}{10}
\providecommand{\url}[1]{#1}
\csname url@samestyle\endcsname
\providecommand{\newblock}{\relax}
\providecommand{\bibinfo}[2]{#2}
\providecommand{\BIBentrySTDinterwordspacing}{\spaceskip=0pt\relax}
\providecommand{\BIBentryALTinterwordstretchfactor}{4}
\providecommand{\BIBentryALTinterwordspacing}{\spaceskip=\fontdimen2\font plus
\BIBentryALTinterwordstretchfactor\fontdimen3\font minus
  \fontdimen4\font\relax}
\providecommand{\BIBforeignlanguage}[2]{{%
\expandafter\ifx\csname l@#1\endcsname\relax
\typeout{** WARNING: IEEEtran.bst: No hyphenation pattern has been}%
\typeout{** loaded for the language `#1'. Using the pattern for}%
\typeout{** the default language instead.}%
\else
\language=\csname l@#1\endcsname
\fi
#2}}
\providecommand{\BIBdecl}{\relax}
\BIBdecl

\bibitem{hamdaoui2007cross-b}
B.~Hamdaoui and P.~Ramanathan, ``Cross-layer optimized conditions for {QoS}
  support in multi-hop wireless networks with {MIMO} links,'' \emph{Selected
  Areas in Communications, IEEE Journal on}, vol.~25, no.~4, 2007.

\bibitem{hamdaoui2007characterization}
B.~Hamdaoui and K.~G. Shin, ``Characterization and analysis of multi-hop
  wireless mimo network throughput,'' in \emph{Proceedings of the 8th ACM
  international symposium on Mobile ad hoc networking and computing}.\hskip 1em
  plus 0.5em minus 0.4em\relax ACM, 2007, pp. 120--129.

\bibitem{su2008cooperative}
W.~Su, A.~K. Sadek, and K.~R. Liu, ``Cooperative communication protocols in
  wireless networks: performance analysis and optimum power allocation,''
  \emph{Wireless Personal Communications}, vol.~44, no.~2, pp. 181--217, 2008.

\bibitem{chakchouk2012uplink}
N.~Chakchouk and B.~Hamdaoui, ``Uplink performance characterization and
  analysis of two-tier femtocell networks,'' \emph{Vehicular Technology, IEEE
  Transactions on}, vol.~61, no.~9, pp. 4057--4068, 2012.

\bibitem{venkatraman2010opportunistic}
P.~Venkatraman, B.~Hamdaoui, and M.~Guizani, ``Opportunistic bandwidth sharing
  through reinforcement learning,'' \emph{Vehicular Technology, IEEE
  Transactions on}, vol.~59, no.~6, pp. 3148--3153, 2010.

\bibitem{dechene2014energy}
D.~Dechene and A.~Shami, ``Energy-aware resource allocation strategies for lte
  uplink with synchronous harq constraints,'' \emph{IEEE Transactions on Mobile
  Computing}, no.~99, pp. 1--1, 2014.

\bibitem{index2017global}
C.~V.~N. Index, ``Global mobile data traffic forecast update, 2016-2021,''
  \emph{Cisco white paper}, 2017.

\bibitem{ahlgren2012survey}
B.~Ahlgren, C.~Dannewitz, C.~Imbrenda, D.~Kutscher, and B.~Ohlman, ``A survey
  of information-centric networking,'' \emph{IEEE Communications Magazine},
  vol.~50, no.~7, 2012.

\bibitem{fayazbakhsh2013less}
S.~K. Fayazbakhsh, Y.~Lin, A.~Tootoonchian, A.~Ghodsi, T.~Koponen, B.~Maggs,
  K.~Ng, V.~Sekar, and S.~Shenker, ``Less pain, most of the gain: Incrementally
  deployable icn,'' in \emph{ACM SIGCOMM Computer Communication Review},
  vol.~43, no.~4.\hskip 1em plus 0.5em minus 0.4em\relax ACM, 2013, pp.
  147--158.

\bibitem{xylomenos2014survey}
G.~Xylomenos, C.~N. Ververidis, V.~A. Siris, N.~Fotiou, C.~Tsilopoulos,
  X.~Vasilakos, K.~V. Katsaros, and G.~C. Polyzos, ``A survey of
  information-centric networking research,'' \emph{IEEE Communications Surveys
  \& Tutorials}, vol.~16, no.~2, pp. 1024--1049, 2014.

\bibitem{NFD:14}
L.~Zhang, A.~Afanasyev, J.~Burke, V.~Jacobson, k.~claffy, P.~Crowley,
  C.~Papadopoulos, L.~Wang, and B.~Zhang, ``Named data networking,''
  \emph{SIGCOMM Comput. Commun. Rev.}, vol.~44, no.~3, pp. 66--73, Jul. 2014.

\bibitem{rossi2011caching}
D.~Rossi and G.~Rossini, ``Caching performance of content centric networks
  under multi-path routing (and more),'' \emph{Relat{\'o}rio t{\'e}cnico,
  Telecom ParisTech}, 2011.

\bibitem{retal2017content}
S.~Retal, M.~Bagaa, T.~Taleb, and H.~Flinck, ``Content delivery network
  slicing: Qoe and cost awareness,'' in \emph{Proc. IEEE ICC}, 2017.

\bibitem{bernardini2013mpc}
C.~Bernardini, T.~Silverston, and O.~Festor, ``Mpc: Popularity-based caching
  strategy for content centric networks,'' in \emph{Proc. of IEEE International
  Conference on Communications (ICC)}, 2013, pp. 3619--3623.

\bibitem{zhang2015survey}
M.~Zhang, H.~Luo, and H.~Zhang, ``A survey of caching mechanisms in
  information-centric networking,'' \emph{IEEE Communications Surveys \&
  Tutorials}, vol.~17, no.~3, pp. 1473--1499, 2015.

\bibitem{ioannou2016survey}
A.~Ioannou and S.~Weber, ``A survey of caching policies and forwarding
  mechanisms in information-centric networking,'' \emph{IEEE Communications
  Surveys \& Tutorials}, vol.~18, no.~4, pp. 2847--2886, 2016.

\bibitem{thar2015efficient}
K.~Thar, T.~Z. Oo, C.~Pham, S.~Ullah, D.~H. Lee, and C.~S. Hong, ``Efficient
  forwarding and popularity based caching for content centric network,'' in
  \emph{Information Networking (ICOIN), 2015 International Conference
  on}.\hskip 1em plus 0.5em minus 0.4em\relax IEEE, 2015, pp. 330--335.

\bibitem{taleb2015user}
T.~Taleb, M.~Bagaa, and A.~Ksentini, ``User mobility-aware virtual network
  function placement for virtual 5g network infrastructure,'' in
  \emph{Communications (ICC), 2015 IEEE International Conference on}.\hskip 1em
  plus 0.5em minus 0.4em\relax IEEE, 2015, pp. 3879--3884.

\bibitem{bagaa2014service}
M.~Bagaa, T.~Taleb, and A.~Ksentini, ``Service-aware network function placement
  for efficient traffic handling in carrier cloud,'' in \emph{Wireless
  Communications and Networking Conference (WCNC), 2014 IEEE}.\hskip 1em plus
  0.5em minus 0.4em\relax IEEE, 2014, pp. 2402--2407.

\bibitem{bernardini2014socially}
C.~Bernardini, T.~Silverston, and O.~Festor, ``Socially-aware caching strategy
  for content centric networking,'' in \emph{Proc. of IEEE Networking
  Conference}.\hskip 1em plus 0.5em minus 0.4em\relax IEEE, 2014, pp. 1--9.

\bibitem{yanqing2016socially}
C.~Yanqing, W.~Muqing, Z.~Min, and W.~Kaili, ``Socially-aware noderank-based
  caching strategy for content-centric networking,'' in \emph{Proc. of IEEE
  International Symposium on Wireless Communication Systems (ISWCS)}, 2016, pp.
  297--303.

\bibitem{ming2014age}
Z.~Ming, M.~Xu, and D.~Wang, ``Age-based cooperative caching in
  information-centric networking,'' in \emph{Proc. of IEEE International
  Conference on Computer Communication and Networks (ICCCN)}, 2014, pp. 1--8.

\bibitem{wang2013collaborative}
S.~Wang, J.~Bi, and J.~Wu, ``Collaborative caching based on hash-routing for
  information-centric networking,'' in \emph{Proc. of ACM SIGCOMM Computer
  Communication Review}, vol.~43, no.~4, 2013, pp. 535--536.

\bibitem{wang2013intra}
J.~M. Wang, J.~Zhang, and B.~Bensaou, ``Intra-as cooperative caching for
  content-centric networks,'' in \emph{Proc of ACM SIGCOMM workshop on
  Information-centric networking}.\hskip 1em plus 0.5em minus 0.4em\relax ACM,
  2013, pp. 61--66.

\bibitem{li2012popularity}
J.~Li, H.~Wu, B.~Liu, J.~Lu, Y.~Wang, X.~Wang, Y.~Zhang, and L.~Dong,
  ``Popularity-driven coordinated caching in named data networking,'' in
  \emph{Proceedings of the eighth ACM/IEEE symposium on Architectures for
  networking and communications systems}.\hskip 1em plus 0.5em minus
  0.4em\relax ACM, 2012, pp. 15--26.

\bibitem{367}
\BIBentryALTinterwordspacing
A.~Afanasyev, I.~Moiseenko, and L.~Zhang, ``{ndnSIM}: {NDN} simulator for
  {NS-3},'' NDN, Technical Report NDN-0005, October 2012. [Online]. Available:
  \url{http://named-data.net/techreports.html}
\BIBentrySTDinterwordspacing

\bibitem{399}
S.~Mastorakis, A.~Afanasyev, I.~Moiseenko, and L.~Zhang, ``{ndnSIM 2.0}: A new
  version of the {NDN} simulator for {NS-3},'' NDN, Technical Report NDN-0028,
  January 2015.

\bibitem{engadget:linknyc:2016}
\BIBentryALTinterwordspacing
D.~Hardawar. (2016, Jan.) Linknyc's free gigabit wifi is here, and it is
  glorious. [Online]. Available:
  \url{https://www.engadget.com/2016/01/19/linknyc-gigabit-wifi-hands-on/}
\BIBentrySTDinterwordspacing

\bibitem{gower1969minimum}
J.~C. Gower and G.~Ross, ``Minimum spanning trees and single linkage cluster
  analysis,'' \emph{Applied statistics}, pp. 54--64, 1969.

\bibitem{manen2013prime}
S.~Manen, M.~Guillaumin, and L.~Van~Gool, ``Prime object proposals with
  randomized prim's algorithm,'' in \emph{Proceedings of the IEEE International
  Conference on Computer Vision}, 2013, pp. 2536--2543.

\bibitem{cloudlet}
M.~Satyanarayanan, P.~Bahl, R.~Caceres, and N.~Davies, ``The case for vm-based
  cloudlets in mobile computing,'' \emph{IEEE Pervasive Computing}, vol.~8,
  no.~4, pp. 14--23, Oct 2009.

\bibitem{chen2016privacy}
M.~Chen, Y.~Qian, J.~Chen, K.~Hwang, S.~Mao, and L.~Hu, ``Privacy protection
  and intrusion avoidance for cloudlet-based medical data sharing,'' \emph{IEEE
  Transactions on Cloud Computing}, 2016.

\bibitem{jiang2015energy}
Z.~Jiang and S.~Mao, ``Energy delay tradeoff in cloud offloading for multi-core
  mobile devices,'' \emph{IEEE Access}, vol.~3, pp. 2306--2316, 2015.

\bibitem{chen2015emc}
M.~Chen, Y.~Zhang, Y.~Li, S.~Mao, and V.~C. Leung, ``Emc: Emotion-aware mobile
  cloud computing in 5g,'' \emph{IEEE Network}, vol.~29, no.~2, pp. 32--38,
  2015.

\bibitem{xu2013survey}
Y.~Xu and S.~Mao, ``A survey of mobile cloud computing for rich media
  applications,'' \emph{IEEE Wireless Communications}, vol.~20, no.~3, pp.
  46--53, 2013.

\bibitem{hassan:15}
H.~Sinky, B.~Hamdaoui, and M.~Guizani, ``Handoff-aware cross-layer assisted
  multi-path {TCP} for proactive congestion control in mobile heterogeneous
  wireless networks,'' in \emph{{GLOBECOM} 2015}, pp. 1--7.

\bibitem{sinky2013cross}
H.~Sinky and B.~Hamdaoui, ``Cross-layer assisted tcp for seamless handoff in
  heterogeneous mobile wireless systems,'' in \emph{Globecom Workshops (GC
  Wkshps), 2013 IEEE}.\hskip 1em plus 0.5em minus 0.4em\relax IEEE, 2013, pp.
  4982--4987.

\bibitem{sinky2016cloudlet}
------, ``Cloudlet-aware mobile content delivery in wireless urban
  communication networks,'' in \emph{Global Communications Conference
  (GLOBECOM), 2016 IEEE}.\hskip 1em plus 0.5em minus 0.4em\relax IEEE, 2016,
  pp. 1--7.

\bibitem{al2004performance}
H.~Al-Zoubi, A.~Milenkovic, and M.~Milenkovic, ``Performance evaluation of
  cache replacement policies for the {SPEC} {CPU2000} benchmark suite,'' in
  \emph{Proceedings of the 42nd annual Southeast regional conference}.\hskip
  1em plus 0.5em minus 0.4em\relax ACM, 2004, pp. 267--272.

\bibitem{lee2001lrfu}
D.~Lee, J.~Choi, J.-H. Kim, S.~H. Noh, S.~L. Min, Y.~Cho, and C.~S. Kim,
  ``{LRFU:} a spectrum of policies that subsumes the least recently used and
  least frequently used policies,'' \emph{IEEE transactions on Computers},
  vol.~50, no.~12, pp. 1352--1361, 2001.

\bibitem{Sarrar:zipf:2012}
\BIBentryALTinterwordspacing
N.~Sarrar, S.~Uhlig, A.~Feldmann, R.~Sherwood, and X.~Huang, ``Leveraging
  zipf's law for traffic offloading,'' \emph{SIGCOMM Comput. Commun. Rev.},
  vol.~42, no.~1, pp. 16--22, Jan. 2012. [Online]. Available:
  \url{http://doi.acm.org/10.1145/2096149.2096152}
\BIBentrySTDinterwordspacing

\bibitem{Faloutsos:zipf:1999}
\BIBentryALTinterwordspacing
M.~Faloutsos, P.~Faloutsos, and C.~Faloutsos, ``On power-law relationships of
  the internet topology,'' \emph{SIGCOMM Comput. Commun. Rev.}, vol.~29, no.~4,
  pp. 251--262, Aug. 1999. [Online]. Available:
  \url{http://doi.acm.org/10.1145/316194.316229}
\BIBentrySTDinterwordspacing

\bibitem{Adamic:zipf:2004}
\BIBentryALTinterwordspacing
L.~A. Adamic and B.~A. Huberman, ``{Zipf's law and the Internet},''
  \emph{Glottometrics}, vol.~3, pp. 143--150, 2002. [Online]. Available:
  \url{http://www.hpl.hp.com/research/idl/papers/ranking/adamicglottometrics.pdf}
\BIBentrySTDinterwordspacing

\bibitem{dohler20165g}
A.~Osseiran, J.~F. Monserrat, and P.~Marsch, \emph{5G Mobile and Wireless
  Communications Technology}.\hskip 1em plus 0.5em minus 0.4em\relax Cambridge
  University Press, 2016.

\end{thebibliography}

\end{document}